\documentclass[prd,preprint,nofootinbib,12pt]{revtex4}
\usepackage{amsfonts}
\usepackage{amsmath}
\usepackage{amssymb}
\usepackage{graphicx}
\usepackage[T1]{fontenc}
\usepackage{xcolor}

\newcommand{\dd}{{\rm{d}}} 

\newcommand{\im}{\mathrm{i}}

\newcommand{\crpartial}{\textup{\rmfamily\dh}}
\newcommand{\C}{{\cal{C}}}

\begin{document}

\title{Static black holes in an external uniform electromagnetic field:
Reissner--Nordstr\"om accelerating in Bertotti--Robinson}

\author{Hryhorii Ovcharenko}
\email{hryhorii.ovcharenko@matfyz.cuni.cz}
\affiliation{Charles University, Faculty of Mathematics and Physics,
Institute of Theoretical Physics,
V~Hole\v{s}ovi\v{c}k\'ach 2, 18000 Prague 8, Czechia}

\author{Ji\v{r}\'i Podolsk\'{y}}
\email{jiri.podolsky@matfyz.cuni.cz}
\affiliation{Charles University, Faculty of Mathematics and Physics,
Institute of Theoretical Physics,
V~Hole\v{s}ovi\v{c}k\'ach 2, 18000 Prague 8, Czechia}

\date{\today}
\begin{abstract}
    We provide a detailed analysis of the non-twisting subcase of the large class of type D black holes with a non-aligned electromagnetic field, presented recently in [H. Ovcharenko and J. Podolsk\'y, Phys. Rev. D {\bf 112} (2025) 064076]. We show that such exact solutions split into two main subclasses that (after a suitable re-parametrization) can be interpreted as either the uncharged Schwarzschild or C--metric in the external Bertotti--Robinson  (BR) spacetime with geometry ${\mathrm{AdS}_2\times\mathrm{S}_2}$, or as the charged Reissner--Nordstr\"{o}m black hole accelerating in the external BR electromagnetic field. The distinction between these two subclasses is determined by the parameter $r_0$ that encodes relations between the external Maxwell field (given by the non-aligned components of the Faraday tensor ${\Phi_0=\Phi_2}$) and the Maxwell field created by the charge of the black hole (given by the aligned component $\Phi_1$). Namely, if ${r_0=0}$ then the electromagnetic field is fully determined by ${\Phi_0=\Phi_2}$, and one gets the C-metric in the BR universe (including also the non-accelerating Schwarzschild--BR black hole). But if ${r_0\neq 0}$ then the electromagnetic field is independently determined by both the external BR field and the field of a black hole itself, and this can be interpreted as the Reissner--Nordstr\"{o}m black hole accelerating in the Bertotti--Robinson spacetime. Even though such an interpretation of the spacetime family is quite simple, it contains a lot of subtleties (e.g. the no-charge limit of the RN--BR spacetime, the non-trivial dependence on the signs of the mass and charge of a black hole, extreme black holes, and others) which we carefully investigate in this work. We also show the explicit relation to solutions previously found by Van den Bergh and Carminati, and we discuss the connection to the Alekseev--Garcia and Alexeev solutions.
\end{abstract}
\maketitle

\tableofcontents

\section{Introduction}
Investigation of axially symmetric (electro-)vacuum spacetimes, possibly with a cosmological constant $\Lambda$, always attracted significant attention because such a class contains the well-know Kerr--Newman family and its important subcases (such as the Schwarzschild, Kerr and Reissner--Nordstr\"{o}m black holes), widely used for astrophysical applications. For the review of these spacetimes and other axially symmetric black holes see e.g. \cite{GriffithsPodolsky:2009,Stephani2003}. Physically, this class can be interpreted as rotating black holes with charge. In fact, according to the uniqueness theorem, the Kerr--Newman spacetime is the most general electrovacuum asymptotically flat spacetime. Among its further important properties let us also recall that it is of algebraic type D, the eigendirections of the Faraday tensor are doubly-aligned with the two principal null directions of the Weyl tensor, and it possesses the Killing tensor \cite{Frolov2017} that allows for the separation of various field equations in these spacetimes \cite{Carter1968,Teukolsky1973}.

Since the discovery of this class of exact solutions, there was a great interest in its generalization by reducing the requirement of asymptotical flatness. Even though it  does not then describe realistic black holes, it is still worth considering such a setup because it may incorporate non-trivial gravitational and electromagnetic fields that describe astrophysically important situations near the black hole itself (as in the case of spacetimes considered in this work, namely the ones immersed in the external electromagnetic field of the Bertotti--Robinson type).

The first significant progress along this direction was the discovery of the Pleba\'nski--Demia\'nski class of solutions \cite{Plebanski1975,Plebanski1976}. It keeps all the algebraic properties of the Kerr--Newman solution (it is of type D and the electromagnetic field is doubly aligned), but instead of the Killing tensor it possesses only the conformal Killing tensor. A detailed physical interpretation was subsequently given in \cite{GriffithsPodolsky:2005,GriffithsPodolsky:2006,PodolskyGriffiths:2006}
(see also the recent developments \cite{Vratny2021,Vratny2023,Ovcharenko2025a,Ovcharenko2025b}), where it was shown that this spacetime, in addition to mass $m$, Kerr parameter $a$, electric and magnetic charges $e$ and $g$, and any cosmological constant $\Lambda$, contains two additional parameters, namely $\alpha$ and~$l$. These parameters have an interpretation as representing  tension and rotation of the string attached to a black hole along the axis of symmetry, causing its acceleration. This string is a particular  topological singularity, and its presence is the only reason why the asymptotical flatness is violated (for  physical analysis of the effects caused by the string, and its influence on the thermodynamics, see e.g. \cite{GriffithsPodolsky:2005,GriffithsPodolsky:2006,Griffiths2006_2,Anabalon2018,Anabalon2019,Liu2022,KolarKrtousOssowski2025}).

Another direction of generalizations of the Kerr--Newman spacetime has been related to the development of various generating techniques. It was shown for the axially symmetric spacetimes that the Ernst equations remain invariant under a wide set of transformations (for details on this and additional information see Part IV in \cite{Stephani2003}). These include the Ehlers, Harrison, and B\"{a}cklund transformations. Among the wide plethora of solutions that can be generated by these methods, we wish to emphasize those generated by the Harrison transformation. This transformation applied to Minkowski spacetime generates the Bonnor--Melvin magnetic universe \cite{Bonnor1954,Melvin1964}, and thus the application of this method to a generic black hole (such as the Kerr--Newman) generates a black hole immersed in the Bonnor--Melvin spacetime \cite{Ernst1976_1,Ernst1976_2}. Such spacetimes are of algebraic type~I (unlike the original Kerr--Newman spacetime of type~D ). Moreover, as one goes away from the black hole the magnetic field becomes weaker while the gravitational field becomes stronger ---  creating some sort of a
``confinement box'' \cite{Melvin1966}. This box, for example, prevents geodesics from escaping to infinity \cite{Galtsov1976,Bizyaev2024,Dadhich1979}. Thus, in this case the asymptotical non-flatness is caused by the external magnetic field of the Bonnor--Melvin type that, even though it has been widely used for the description of various astrophysical effects near black holes, suffers from the problems related to the existence of such an unphysical confinement box.

Recently, a physically interesting generalization of the famous family of black holes appeared that assumes that the resulting spacetimes remain of algebraic type~D, but does not require the alignment condition for the Maxwell field. The main reason why this setup was not considered previously is because the corresponding field equations are complicated to solve. A significant step in this direction was the discovery of the solution \cite{VandenBergh2020} by Van den Bergh and Carminati. In this work the authors found the Robinson--Trautman  (i.e. shear-free, non-twisting, and expanding) solution of algebraic type D with a non-aligned electromagnetic field. Unfortunately, the work \cite{VandenBergh2020} is purely mathematical and does not provide a physical analysis of this class of solutions.

This motivated us to (i) generalize this solution to the case with a non-zero twist, (ii) give a physical interpretation to the resulting spacetimes. Both these questions were partially answered in our recent works \cite{Ovcharenko2025,Podolsky2025}. In particular, we have found and presented a generalization of the Van den Bergh--Carminati solution to the twisting case (thus going beyond the Robinson--Trautman family) and we discovered that (if the black hole has no charge) the non-aligned component of the Faraday tensor represents the external uniform electromagnetic field of the Bertotti--Robinson type --- which is the physical reason for asymptotical non-flatness in this case. Such a new setting is \emph{different} from the Melvin-based spacetimes and possesses some more realistic properties because the external field is asymptotically uniform and does not create an infinitely strong gravitational field at infinity. Such favorite properties already allowed several researchers to use this spacetime as a playground for studying various physical effects near black holes, namely magnetic reconnection and Penrose effect \cite{Zeng2025,Mirkhaydarov2025}, light deflection with black hole shadows \cite{Wang2025,Zeng2025_2,Ali2025}, motion of particles and various field perturbations \cite{Gray2025,Zhang2025}, and others. Interestingly, also a class of exact solutions was recently found \cite{Astorino2025} which represent black holes in the \emph{combined} Bertotti--Robinson and Bonnor--Melvin external electromagnetic fields.

However, there are still many ``blind spots'' within this large class of solutions, for example to explicitly clarify the relation of the static subcase of \cite{Ovcharenko2025} to the Van den Bergh--Carminati solution \cite{VandenBergh2020}. Also, the physical interpretation of the \emph{charged} black holes is still unclear. This is caused by the fact that the original solution \cite{Ovcharenko2025} contains many integration constants that have an interpretation of charge and the external field only in some special subcases, while their interaction in the strong-field regime is obscured. Our current work aims to cover some of these issues and give a better physical interpretation to the non-twisting subcase of \cite{Ovcharenko2025}, also relating it to the Van den Bergh and Carminati solution \cite{VandenBergh2020}. Another open problem is to find the relation of our class of spacetimes \cite{Ovcharenko2025} to the Alekseev--Garcia \cite{Alekseev1996}  and Alekseev \cite{Alekseev2025} solutions.

We structure our work in the following way. In Sec.~\ref{sec:review} we review our new general class of type~D solutions with non-aligned  electromagnetic field, and in Sec.~\ref{section-no twist} we systematically derive its complete non-twisting limit. Sec.~\ref{sec:alter} introduces another convenient form of this class of black holes, and discusses in detail all possible subcases, both uncharged and charged. Explicit relation to the Van den Bergh--Carminati solution, and several comments on the Alekseev--Garcia and Alekseev solutions, are contained in Sec.~\ref{section-VdBC} and Sec.~\ref{section-Alexeev}, respectively. Summary of our results and some concluding remarks are given in the final Sec.~\ref{sec:final}.

\newpage

\section{Review of the new class}
\label{sec:review}

Recently, we found a new class of type~D solutions to the Einstein--Maxwell equations~\cite{Ovcharenko2025} which describe  various black holes in an external electromagnetic field that is \emph{not aligned} with the Weyl tensor structure. In particular, it contains a novel interesting exact Kerr--BR spacetime \cite{Podolsky2025} such that the Kerr black hole is immersed in an external magnetic (or electric) field of the Bertotti--Robinson form.

This large class is generally described by the metric
\begin{align}
    \dd s^2 =\dfrac{1}{\Omega^2}\Big[&-\dfrac{\mathcal{Q}}{\varrho^2}\big[(\dd\tau-\omega\,(x+x_0)^2 \dd\phi\big]^2
    +\dfrac{\varrho^2}{\mathcal{Q}}\,\dd r^2\nonumber\\
   &+\dfrac{\mathcal{P}}{\varrho^2}\big[\,\omega\,\dd\tau+(r+r_0)^2\dd\phi\big]^2
    +\dfrac{\varrho^2}{\mathcal{P}}\,\dd x^2 \,\Big]\,,\label{ds_alpha_omega}
\end{align}
where the functions are
\begin{align}
  \varrho^2 &=(r+r_0)^2+\omega^2(x+x_0)^2\,,       \label{rho2_rescl}  \\
  \Omega^2  &= (1-\alpha rx)^2 +2\alpha^2\C(r^2-\omega^2x^2)+\alpha^2 \Big(4\alpha^2\omega^2\,\C^2-2\C\dfrac{\epsilon}{k}\Big)\,r^2x^2 \nonumber\\
         & \hspace{18mm}+\dfrac{4\alpha^2\C}{(1+\alpha^2\omega^2\,\C^2)k}
         \Big[(m+\alpha \omega\,n\, \C)\,x + (n-\alpha \omega\, m\, \C)\dfrac{r}{\omega}\,\Big]\,rx\,, \label{Om2_rescl}
\end{align}
and $\mathcal{P}, \mathcal{Q}$ are quartic polynomials of $x$ and $r$, respectively,
\begin{align}
    \mathcal{P}(x)&= a_0+a_1\,x+a_2\,x^2+a_3\,x^3+a_4\,x^4\,,\label{P_prime}\\
    \mathcal{Q}(r)&= b_0+b_1\,r+b_2\,r^2+b_3\,r^3+b_4\,r^4\,.\label{Q_prime}
\end{align}
The coefficients $a_i$ and $b_i$ are explicitly given by
\begin{align}
    a_0&= k\,, \nonumber\\
    a_1&= \frac{2}{\omega}\,\dfrac{n-\alpha\omega\, m\,\C}{1+\alpha^2\omega^2\,\C^2}\,, \nonumber\\
    a_2&=-\epsilon \,,\label{prim_param_6}\\
    a_3&= 2\alpha\,\dfrac{(1+2\alpha^2\omega^2\,\C^2)m-\alpha\omega\, n\,\C}{1+\alpha^2\omega^2\,\C^2}\,, \nonumber\\
    a_4&=-\alpha^2\omega^2(1+4\alpha^2\omega^2\,\C^2)k + 2\alpha^2\omega^2\C \Big(\epsilon-\dfrac{(m+\alpha\omega\, n\,\C)^2}{\omega^2k(1+\alpha^2\omega^2\,\C^2)^2}\Big)\,, \nonumber
\end{align}
and
\begin{align}
    b_0&= \omega^2 k\,, \nonumber\\
    b_1&=-2\,\dfrac{m+\alpha\omega\,n\,\C}{1+\alpha^2\omega^2\,\C^2}\,, \nonumber\\
    b_2&= \epsilon \,,\label{prim_param_1}\\
    b_3&=-2\,\frac{\alpha}{\omega}\,\dfrac{(1+2\alpha^2\omega^2\,\C^2)n+\alpha \omega\, m\,\C}{1+\alpha^2\omega^2\,\C^2}\,, \nonumber\\
    b_4&=-\alpha^2(1+4\alpha^2\omega^2\,\C^2)k + 2\alpha^2\C\Big(\epsilon+\dfrac{(n-\alpha\omega\, m\,\C)^2}{\omega^2k(1+\alpha^2\omega^2\,\C^2)^2}\Big)\,. \nonumber
\end{align}
The real constant $\C$ is a useful shorthand
\begin{equation}
  \C := 2\,c\,\bar{c}\, k \,,
    \label{C definition}
\end{equation}
in which $c$ is a complex parameter, while $k, \alpha, \omega, m, n, \epsilon$ are real parameters.
This gravitational field is coupled to the Maxwell field whose electromagnetic 4-potential is
\begin{align}
    \mathbf{A} = \dfrac{1}{4\alpha\,\bar{c}}\Big[\,
    \dfrac{ \omega\,\Omega_{,r}-\im\,\Omega_{,x}}{(r+r_0)+\im\,\omega\,(x+x_0)}\,\dd \tau
    +\Big(\dfrac{(r+r_0)^2\,\Omega_{,r}+\im\,\omega\,(x+x_0)^2\,\Omega_{,x}}{(r+r_0)+\im\,\omega\,(x+x_0)}-\Omega\Big)\,\dd\phi
    \Big].   \label{A-3}
\end{align}
It contains the complex constant~$c$, which can naturally be parametrized by its absolute value $|c|$ and the (duality) rotation parameter $\gamma$ as
\begin{align}
    c=|c|\,e^{\im\, \gamma} \,.   \label{def-gamma}
\end{align}

Let us also recall that the additional two real coefficients $r_0$ and $x_0$ are not mutually independent. They are constrained by the relation
\begin{align}
  (\omega\,x_0-A)^2  + (\,r_0-B)^2 = R^2\,,\label{r0x0_cond}
\end{align}
where the constants $A, B, R$ are
\begin{align}
A &= \dfrac{2\C}
{(1+\alpha^2\omega^2\,\C^2)[(1+4\alpha^2\omega^2\,\C^2)k - 2\C\epsilon]}\,(n-\alpha\omega\,m\,\C)\,,\nonumber\\
B &= \dfrac{2\C}
{(1+\alpha^2\omega^2\,\C^2)[(1+4\alpha^2\omega^2\,\C^2)k - 2\C\epsilon]}\,(m+\alpha\omega\,n\,\C)\,,\label{defABnew}\\
R &= \dfrac{2\C}
{\sqrt{1+\alpha^2\omega^2\,\C^2}\,[(1+4\alpha^2\omega^2\,\C^2)k - 2\C\epsilon]}\,\sqrt{m^2+n^2}\,.\nonumber
\end{align}
The constraint \eqref{r0x0_cond} is automatically satisfied by introducing a single angular parameter $\beta$ such that
\begin{align}
 \omega\,x_0 &= A + R \sin\beta\,,\nonumber\\
         r_0 &= B + R\cos\beta\,.\label{x0r0-in-ABR}
\end{align}
It means that $\omega\,x_0$ and $r_0$ are fully determined by the remaining parameters. In particular, notice that \emph{both} the coefficients $\omega\,x_0, r_0$ can \emph{always be made zero} (${\omega\,x_0=0=r_0}$) by the special choice of the duality parameter $\beta$ such that
\begin{align}
    \tan \beta_0=\dfrac{n\,-\alpha\omega\,m\,\C}{m+\alpha\omega\,n\,\C}\,. \label{beta_cond-new}
\end{align}

In our investigations here we will employ the metric form \eqref{ds_alpha_omega}--\eqref{C definition}
and \eqref{A-3}, instead of the other equivalent parameterizations presented in \cite{Ovcharenko2025}, because it is more convenient for obtaining the static limit. Let us summarize that it contains 9 free real parameters, namely $\alpha, \omega, k, m, n, \epsilon, \beta, \gamma, |c|$. In the context of the present paper it is important to recall that $\omega$ is the parameter representing the \emph{twist} of the (null, geodesic, shearfree) congruence generated by the two principal null directions of the Weyl tensor \cite{Ovcharenko2025}. The physical meaning of the remaining parameters will be clarified below.

For these studies, the optical, Weyl, and Faraday scalars are important. They are expressed with respect to a null frame ${(\mathbf{k}, \mathbf{l}, \mathbf{m},  \bar{\mathbf{m}})}$ such that ${\mathbf{k} \cdot \mathbf{l}=-1}$ and ${\mathbf{m} \cdot \bar{\mathbf{m}}=1}$. It is most natural to adapt it along the two (double-degenerate) principal null directions (PNDs) of the Weyl tensor of type~D. Such a frame is explicitly given by
\begin{align}
    \mathbf{k}&=\sqrt{\dfrac{\Omega^2}{2\mathcal{Q}\varrho^2}}\,
      (\,r^2\partial_{\tau}-\omega \partial_{\phi}+\mathcal{Q}\,\partial_r)\,,\nonumber\\
    \mathbf{l}&=\sqrt{\dfrac{\Omega^2}{2\mathcal{Q}\varrho^2}}\,
      (\,r^2\partial_{\tau}-\omega\partial_{\phi}-\mathcal{Q}\,\partial_r)\,,\label{null_tetr}\\
    \mathbf{m}&=\sqrt{\dfrac{\Omega^2}{2\mathcal{P}\varrho^2}}\,(\,\omega x^2\partial_{\tau}+\partial_{\phi}+\im\, \mathcal{P}\,\partial_x) \,,\nonumber
\end{align}
see \cite{Ovcharenko2025}. The corresponding optical scalars are
\begin{align}
    \kappa&=\nu=0,\qquad\sigma=\lambda=0\,,\nonumber\\
    \rho_{\rm sc}&=\mu=\dfrac{1}{2}\sqrt{\dfrac{\mathcal{Q}\, \Omega^2}{2\varrho^2}}\Bigg[\Big(\ln \dfrac{\Omega^2}{\varrho^2}\Big)_{,r}+\dfrac{\im}{\omega}\, (\ln\varrho^2)_{,x}\Bigg],\nonumber\\
    \tau&=\pi=\dfrac{1}{2}\sqrt{\dfrac{\mathcal{P}\,\Omega^2}{2\varrho^2}}\Bigg[\,\omega \,(\ln \rho^2)_{,r}+\im\, \Big(\ln \dfrac{\Omega^2}{\varrho^2}\Big)_{,x}\Bigg],\label{opt_sc_5}\\
    \alpha&=\beta=\dfrac{1}{4}\sqrt{\dfrac{\mathcal{P}\,\Omega^2}{2\varrho^2}}\Bigg[\,\omega\,(\ln \varrho^2)_{,r}+\im\, \Big(\ln \dfrac{\mathcal{P}}{\varrho^2\Omega^2}\Big)_{,x}\Bigg],\nonumber\\
    \epsilon&=\gamma=\dfrac{1}{4}\sqrt{\dfrac{\mathcal{Q}\,\Omega^2}{2\varrho^2}}\Bigg[ \Big(\ln \dfrac{\mathcal{Q}}{\varrho^2\Omega^2}\Big)_{,r}+\dfrac{\im}{\omega}\,(\ln \varrho^2)_{,x}\Bigg].\nonumber
\end{align}
(It may seem that these expressions diverge in the limit ${\omega\to 0}$. However, as we will show in the next section, it is not so. The corresponding limit is actually finite.)
With respect to the frame \eqref{null_tetr} the only non-zero Weyl scalar is
\begin{align} \label{psi_rescaled}
    \Psi_2=&\dfrac{\Omega^2}{12}\dfrac{(\,r+r_0+\im\,\omega\,(x+x_0))^2}{r+r_0-\im\,\omega \,(x+x_0)}\nonumber\\
    &\times\Bigg[\Big(\dfrac{\mathcal{Q}}{(r+r_0+\im\,\omega\,(x+x_0))^3}\Big)_{,rr}
    +\Big(\dfrac{\mathcal{P}}{(r+r_0+\im\,\omega\,(x+x_0))^3}\Big)_{,xx}\Bigg],
\end{align}
and the electromagnetic scalars representing the Maxwell field are
\begin{align}
\Phi_0&=\Phi_2=\dfrac{\alpha |c|e^{\im\gamma}}{\Omega}\,\dfrac{\sqrt{\mathcal{P}\mathcal{Q}}}{(r+r_0)+\im\,\omega\,(x+x_0)}\,,
    \label{phi_rescaled} \\
\Phi_1&=\dfrac{e^{\im\gamma}}{4\alpha\,|c|}\,\dfrac{\Omega^2}{[(r+r_0)+\im\,\omega\,(x+x_0)]^2}\,\Big[\,
     \omega\,(x+x_0)^2\Big(\dfrac{\Omega_{,r}}{x+x_0}\Big)_{,x}
     -\im\,  (r+r_0)^2\Big(\dfrac{\Omega_{,x}}{r+r_0}\Big)_{,r} \,\Big].\nonumber
\end{align}

\section{Deriving the non-twisting subcase}
\label{section-no twist}

The purpose of our work is to systematically  analyze \emph{all possible static solutions} contained in our new large class of spacetimes \eqref{ds_alpha_omega}. These are obtained by setting the twist parameter~$\omega$ to zero in all the above expressions. However, one has to be careful in considering such a limit because some of the terms in the metric functions may diverge for ${\omega\to 0}$.

\subsection{Pleba\'{n}ski--Demia\'{n}ski form of the metric}

We start with the metric \eqref{ds_alpha_omega}--\eqref{C definition}. By inspecting the metric coefficients \eqref{prim_param_6}, \eqref{prim_param_1} it can be seen that for obtaining their reasonable ${\omega \to 0}$ limit it is necessary that $\dfrac{n}{\omega}$ \emph{remains finite}, which requires ${n\to0}$. Introducing the constant
\begin{align}
    n' :=\dfrac{n}{\omega}\,,
\end{align}
and keeping the parameters $\alpha, m, |c|, k, \epsilon$ constant, the coefficients \eqref{prim_param_6} and \eqref{prim_param_1} simplify to
\begin{align}
    a_0&= k\,, \nonumber\\
    a_1&= 2 (n'-2 \alpha m |c|^2 k)\,, \nonumber\\
    a_2&=-\epsilon \,,\label{a-param}\\
    a_3&= 2\alpha m\,, \nonumber\\
    a_4&= -4\alpha^2m^2|c|^2\,, \nonumber
\end{align}
and
\begin{align}
    b_0&= 0\,, \nonumber\\
    b_1&=-2 m\,, \nonumber\\
    b_2&= \epsilon \,,\label{b-param}\\
    b_3&=-2\alpha\,(n'+2\alpha m |c|^2 k)\,, \nonumber\\
    b_4&=-\alpha^2 \big[\, k - 4|c|^2 \big(k\,\epsilon+(n'-2\alpha m |c|^2 k)^2\big)\big]\,. \nonumber
\end{align}
Let us note that the special case ${n=0}$ directly leads to the ${n'=0}$ subcase of \eqref{a-param} and \eqref{b-param}.
The static (i.e. nontwisting) limit is easily obtained by simply setting ${\omega=0}$ (keeping ${x_0=\mathrm{const.}}$) in \eqref{ds_alpha_omega} and \eqref{A-3}, using the fact that now ${\varrho^2=(r+r_0)^2}$. The resulting metric is
\begin{align}
    \dd s^2=\dfrac{1}{\Omega^2}\Big[-\dfrac{\mathcal{Q}}{(r+r_0)^2}\,\dd\tau^2
           +(r+r_0)^2\Big(\dfrac{\dd r^2}{\mathcal{Q}}
           +\dfrac{\dd x^2}{\mathcal{P}}+\mathcal{P}\,\dd\phi^2 \Big)\Big],
           \label{metric-nontwist-PD}
           \end{align}
and the corresponding Maxwell field reads
\begin{align}
    \mathbf{A} = \dfrac{1}{4\alpha\,\bar{c}}\,\Big[\,
    \dfrac{ -\im\,\Omega_{,x}}{r+r_0}\,\dd \tau
    +\big((r+r_0)\,\Omega_{,r}-\Omega\big)\dd\phi\, \Big].   \label{A-nontwist}
\end{align}
The conformal factor \eqref{Om2_rescl} becomes
\begin{align}
  \Omega^2(r,x) &= (1-\alpha rx)^2  +4\alpha^2|c|^2\,
       \big[\, 2m\,rx^2 + 2(n'-2\alpha m|c|^2 k)\,r^2 x
         +(k-\epsilon\,x^2)\,r^2\,\big]\,, \label{Om2_stat}
\end{align}
and the quartic metric functions \eqref{P_prime}, \eqref{Q_prime} take the explicit form
\begin{align}
    \mathcal{P}(x)&= k+2 (n'-2 \alpha m |c|^2 k)\,x-\epsilon\,x^2+2\alpha m\,x^3-4\alpha^2m^2|c|^2\,x^4\,,\label{P-nontwist-PD}\\
    \mathcal{Q}(r)&= -2 m\,r+\epsilon\,r^2-2\alpha\,(n'+2\alpha m |c|^2 k)\,r^3 \nonumber\\
    & \quad\,-\alpha^2 \big[\, k - 4|c|^2 \big(k\,\epsilon+(n'-2\alpha\,m |c|^2 k)^2\big)\big]\,r^4\,.\label{Q-nontwist-PD}
\end{align}

Moreover, the constants \eqref{defABnew} reduce to
\begin{align}
A = 0\,,\qquad
B = R = \dfrac{4m|c|^2}{1 - 4\epsilon|c|^2} \,,\label{defABnew-nontwist}
\end{align}
so that the constraint \eqref{r0x0_cond} simplifies to
\begin{align}
 (\,r_0-R)^2 = R^2\,,\label{r0x0_cond-nontwist}
\end{align}
implying ${r_0=R\pm|R|}$, which corresponds to the two special values ${\beta=\pi}$ and ${\beta=0}$. There are thus \emph{two distinct types of solutions}, namely
\begin{align}
    \hbox{ case a)\,:}&\qquad  r_0=0\,,\label{r0-two cases_1}\\
    \hbox{ case b)\,:}&\qquad  r_0=\dfrac{8m|c|^2}{1-4\epsilon|c|^2}\,.
    \label{r0-two cases_2}
\end{align}

We observe from  \eqref{phi_rescaled}  that in the limit ${|c|\to 0}$ the electromagnetic field becomes aligned (it actually vanishes, see below), and in such a limit these two cases coincide, with the same value ${r_0=0}$. The metric then reduces to
\begin{align}
    \dd s^2=\dfrac{1}{(1-\alpha rx)^2}\Big[
       -\dfrac{\mathcal{Q}}{r^2}\,\dd\tau^2 + \dfrac{r^2}{\mathcal{Q}}\,\dd r^2
           +r^2\Big(\dfrac{\dd x^2}{\mathcal{P}}+\mathcal{P}\,\dd\phi^2 \Big)\Big],
           \label{metric-nontwist-PD-c=0}
\end{align}
with
\begin{align}
    \mathcal{P}&= k+2 n'\,x-\epsilon\,x^2+2\alpha m\,x^3\,,\label{P-nontwist-PD-c=0}\\
    \mathcal{Q}&= -2 m\,r+\epsilon\,r^2-2\alpha\,n'\,r^3 -\alpha^2 k\,r^4\,.\label{Q-nontwist-PD-c=0}
\end{align}
This is exactly the \emph{Pleba\'{n}ski--Demia\'{n}ski} metric \cite{Plebanski1976} of type~D, as given by Eqs.~(16.5)--(16.6) with  ${\omega=0}$ and ${\Lambda=0}$ in the monograph~\cite{GriffithsPodolsky:2009}. It is a general \emph{vacuum} solution.

Indeed, as we discussed in \cite{Ovcharenko2025}, the  quantity $r_0$ is related to electric and magnetic charges $e$~and~$g$ of the black hole via the relation ${e^2+g^2=r_0^2/(4|c|^2)}$, see Eq.~(113) therein.  In the first case~a) we immediately get ${e^2 + g^2 = 0}$ because ${r_0 = 0}$ even if ${|c| \neq 0}$. In the second case~b) we also obtain ${e^2 + g^2 \approx 16m^2|c|^2 \to 0}$ in the limit $|c|\to 0$. This leads to vanishing of not only the non-aligned part of the electromagnetic field, but also of its aligned part (related to the black hole charges $e$~and~$g$) in the ${|c|\to 0}$ limit, for \emph{both} the cases a) and b).

For the general class of non-twisting spacetimes \eqref{metric-nontwist-PD}, the null tetrad \eqref{null_tetr} simplifies to
\begin{align}
    \mathbf{k}&=\dfrac{\Omega}{\sqrt{2\mathcal{Q}}\,(r+r_0)}\,
      (\,r^2\partial_{\tau}+\mathcal{Q}\,\partial_r)\,,\nonumber\\
    \mathbf{l}&=\dfrac{\Omega}{\sqrt{2\mathcal{Q}}\,(r+r_0)}\,
      (\,r^2\partial_{\tau}-\mathcal{Q}\,\partial_r)\,,\label{null_tetr_notwist}\\
    \mathbf{m}&=\dfrac{\Omega}{\sqrt{2\mathcal{P}}\,(r+r_0)}\,
    (\,\partial_{\phi}+\im\, \mathcal{P}\,\partial_x) \,,\nonumber
\end{align}
in which the key optical scalars \eqref{opt_sc_5}, using the relation ${\displaystyle\lim_{\omega\to0}\big[\omega^{-1}(\ln\varrho^2)_{,x}\big]  = 0 }$, read
\begin{align}
    \kappa&=\nu=0,\qquad\sigma=\lambda=0\,,\nonumber\\
    \rho_{\rm sc}&=\mu=\sqrt{\dfrac{\mathcal{Q}}{2}}\,
       \Big(\dfrac{\Omega}{r+r_0}\Big)_{,r}\,,\label{opt_sc_6}
\end{align}
confirming that both PNDs of these spacetimes generate geodesic shear-free null congruences, that are twist-free but expanding (because $\rho_{\rm sc}$ and $\mu$ are real but non-zero).

Finally, in the  static limit ${\omega\to0}$ the Weyl scalar \eqref{psi_rescaled} reduces to
\begin{align}\label{psi_notwist}
    \Psi_2=&\dfrac{1}{12}\,\Omega^2\,(r+r_0)\Bigg[
     \Big(\dfrac{\mathcal{Q}}{(r+r_0)^3}\Big)_{,rr}
    +\Big(\dfrac{\mathcal{P}}{(r+r_0)^3}\Big)_{,xx}\Bigg],
\end{align}
while the Maxwell field scalars \eqref{phi_rescaled} simplify considerably to
\begin{align}
\Phi_0&=\Phi_2= \alpha |c|\,\dfrac{e^{\im\gamma}}{\Omega}\,\dfrac{\sqrt{\mathcal{P}\mathcal{Q}}}{r+r_0}\,,
    \nonumber \\
\Phi_1&=\dfrac{-\im\, e^{\im\gamma}}{4\alpha\,|c|}\,\Omega^2\, \Big(\dfrac{\Omega_{,x}}{r+r_0}\Big)_{,r} \,.\label{phi_notwist}
\end{align}

Moreover, it can be seen that in the limit ${|c|\to0}$ we get ${\Phi_0=\Phi_2=0}$ and also ${\Phi_1=0}$. It confirms that the ${|c|=0}$ subcase corresponds to a subcase of the Pleba\'{n}ski--Demia\'{n}ski \emph{vacuum} solution.

\subsection{Griffiths--Podolsk\'{y} form of the metric}

To proceed further with the analysis of the derived metric \eqref{metric-nontwist-PD}, it is convenient to transform it into another form which is a generalization of the Griffiths--Podolsk\'{y} (GP) form \cite{GriffithsPodolsky:2005, GriffithsPodolsky:2006, PodolskyGriffiths:2006, GriffithsPodolsky:2009}  of the Pleba\'{n}ski--Demia\'{n}ski (PD) metric \cite{Plebanski1976}. Its advantage is that it provides a spherical-like coordinate system with much simpler metric functions, in which the poles and horizons are explicitly identified, and the free parameters of the exact solution to the Einstein--Maxwell equations is given a clear physical meaning. The key step is to put the metric function $\mathcal{P}$,  given by the general quartic polynomial~\eqref{P_prime}, to a \emph{factorized form}
\begin{align}\label{P-factorized}
 \mathcal{P}=(1-x^2)(1-a_3\,x-a_4\,x^2)\,.
\end{align}
This imposes the constraints ${a_1+a_3=0}$, ${a_0+a_2+a_4=0}$, and ${a_0=1}$.\footnote{Black hole geometries with non--spherical topologies can also be obtained for ${a_0=-1}$ and ${a_0=0}$.} In view of \eqref{a-param}, such conditions imply the following unique choice of the Pleba\'{n}ski--Demia\'{n}ski parameters:
\begin{align}\label{kneps}
    k=1\,,\qquad
    n'=-\alpha m + 2\alpha m|c|^2\,,\qquad
    \epsilon=1-4\alpha^2m^2|c|^2\,.
\end{align}
The metric function $\mathcal{P}(x)$ thus takes the factorized form \eqref{P-factorized} with
 \begin{align}\label{a3a4}
 a_3=2\alpha m\,,\qquad
 a_4=-4 \alpha^2 m^2 |c|^2\,.
\end{align}
Interestingly, for such a choice  the coefficients \eqref{b-param} become
\begin{align}
    b_1&=-2 m\,, \qquad
    b_2 = 1-4\alpha^2m^2|c|^2 \,,\nonumber\\
    b_3&= -\alpha^2(1 - 4|c|^2)\,b_1 \,, \qquad
    b_4 = -\alpha^2(1 - 4|c|^2)\,b_2 \,, \label{bi-fact}
\end{align}
so that the metric function $\mathcal{Q}(r)$, given by \eqref{Q_prime}, \emph{also factorizes} to
\begin{align}\label{Q-factorized}
\mathcal{Q}= r \,\big(\,b_1+b_2\,r\big)\big[\,1-\alpha^2(1 - 4|c|^2)\,r^2\,\big].
\end{align}

It only remains to introduce the angular coordinate $\theta$ by ${x=\cos\theta}$, so that the metric            \eqref{metric-nontwist-PD} with the factorized metric functions \eqref{P-factorized}, \eqref{Q-factorized}
takes a nice  compact form
\begin{align}\label{metric-nontwist-GP-r0}
    \dd s^2=\dfrac{1}{\Omega^2}\Big[-\dfrac{\mathcal{Q}}{(r+r_0)^2}\,\dd\tau^2
           +(r+r_0)^2\Big(\dfrac{\dd r^2}{\mathcal{Q}}
           +\dfrac{\dd \theta^2}{{P}}+{P}\sin^2\theta \,\dd\phi^2 \Big)\Big],
\end{align}
where, defining ${P}$ via the relation ${\mathcal{P}=(1-x^2)\,{P}}$,
\begin{align}
    {P}& =1-2\alpha m\cos\theta+4\alpha^2m^2 |c|^2\cos^2\theta\,,\label{metric-function-P-nontwist}\\
    \mathcal{Q}& =r\,\big[ (1-4\alpha^2m^2|c|^2)\,r -2m \big]\big[1-\alpha^2(1-4|c|^2)\,r^2\,\big].
    \label{metric-function-Q-nontwist}
\end{align}
With \eqref{kneps}, the conformal factor \eqref{Om2_stat} becomes
\begin{align}\label{Omega-nontwist}
    \Omega^2&=(1-\alpha r\cos\theta)^2+8\alpha^2|c|^2(\cos\theta-\alpha r)\,mr\cos\theta
    \nonumber\\
    &\hspace{20mm} +4\alpha^2|c|^2r^2 ( \sin^2\theta + 4\alpha^2m^2|c|^2 \cos^2\theta)\,.
\end{align}

Recall that the physical parameters are now $m, \alpha, |c|$. Moreover, there is the parameter $r_0$ which \emph{takes just two values} \eqref{r0-two cases_1} or \eqref{r0-two cases_2}, distinguishing the two distinct subclasses, the case a) and the case b).\footnote{By substituting \eqref{kneps} into \eqref{beta_cond-new}, in the limit ${\omega=0}$ we get ${\tan\beta_0=0}$ which has just two solutions, namely ${\beta_0=0}$ and ${\beta_0=\pi}$. They simplify \eqref{defABnew} and \eqref{x0r0-in-ABR} to ${A=0}$, ${|B|=R}$, and ${\omega\,x_0 = 0}$, ${r_0 = B \pm R}$, consistent with \eqref{defABnew-nontwist}. The solution ${\beta_0=\pi}$ gives the case a), while the solution ${\beta_0=0}$  gives the case b).} In addition, there is the duality parameter $\gamma$ of the electromagnetic field, see the expressions \eqref{phi_notwist}.

\subsection{Basic physical interpretation: black hole horizons, axes, and charges}
\label{basic-interpretation}

This is an interesting large family of black holes with spherical topology of horizons.
In particular, for ${\alpha=0}$ (keeping the remaining two parameters constant) we obtain
\begin{align}
    {P}& = 1\,,\qquad
    \mathcal{Q} = r\,(r -2m)\,,\qquad
    \Omega^2=1\,,
\end{align}
so that in the ${r_0=0}$ subcase we recover the \emph{Schwarzschild black hole} in standard spherical coordinates.
For  ${|c|=0}$ (implying ${r_0=0}$) we get
\begin{align}
    {P}& = 1-2\alpha m\cos\theta\,,\qquad
    \mathcal{Q} = r\,(r -2m)(1-\alpha^2r^2)\,,\qquad
    \Omega^2=(1-\alpha r\cos\theta)^2\,,
\end{align}
which is the usual form of the C-metric, see Eqs.~(14.6) and~(14.7) in \cite{GriffithsPodolsky:2009}, that represents uniformly \emph{accelerating} black holes.

Due to the nice factorization \eqref{metric-function-Q-nontwist} of the metric function $\mathcal{Q}$, it is very easy to explicitly determine the position of the horizons, given by the condition ${\mathcal{Q}(r)=0}$, namely at
\begin{align}\label{horiozons-nontwist}
    r_b = \frac{2m}{1-4\alpha^2m^2|c|^2} \,,\qquad
    r_a = \frac{1}{\alpha}\frac{1}{\sqrt{1-4|c|^2}}\,.
\end{align}
The former is the \emph{black-hole horizon}, while the latter is the \emph{acceleration horizon} (here we naturally assume ${r>0}$ and ${\alpha>0}$).

Because of the presence of the factor $\sin^2\theta$ in the metric \eqref{metric-nontwist-GP-r0}, these black holes have \emph{poles} located at ${\theta=0}$ and ${\theta=\pi}$. In general, they are not regular (they contain conical singularities associated with the ``cosmic strings'' or ``struts''). The \emph{regularity of these axes}, with the range of the angular coordinate
\begin{align}\label{range-of-phi}
\phi\in[0,2\pi C)\,,
\end{align}
is given by the condition ${C\,{P}(\theta=0 \hbox{\ or\ } \theta=\pi)=1}$. This can be achieved (separately) by the \emph{specific conicity}
\begin{align}\label{regularity-nontwist}
C = \frac{1}{1\mp 2\alpha m+4\alpha^2m^2 |c|^2}\,.
\end{align}
If (and only if) ${\alpha m =0}$ then \emph{both} the axes are regular.

As was already mentioned in the discussion after Eqs.~\eqref{P-nontwist-PD-c=0}--\eqref{Q-nontwist-PD-c=0}, both the limits ${\alpha=0}$ and ${|c|=0}$ lead to the elimination of the electromagnetic field. Nevertheless, after introduction of a suitable new parameter~$B$, in the next section we will prove that the general family \eqref{metric-nontwist-GP-r0} actually contains charged Reissner--Nordstr\"om black holes, and also (possibly charged and accelerating) black holes in an external Bertotti--Robinson electromagnetic field.

To exactly evaluate the \emph{charges of such black holes}, we have to take the real counterpart of the 1-form $\mathbf{A}$, given in \eqref{A-nontwist}, namely
\begin{align}
    \mathbf{A^{real}}\equiv 2\,\mathrm{Re}(\mathbf{A})=\dfrac{1}{2\alpha |c|}\,\Big[
    \sin\gamma\,\dfrac{\Omega_{,x}}{r+r_0}\,\dd\tau+\cos\gamma\,\big((r+r_0)\Omega_{,r}-\Omega\big)\dd\phi\Big],
\end{align}
then calculate the Faraday 2-form $\mathbf{F}=\dd\mathbf{A}^{\mathrm{real}}$,
\begin{align}\label{F}
    \mathbf{F}&=\dfrac{\sin\gamma}{2\alpha |c|}\,\Big(\dfrac{(r+r_0)\Omega_{,rx}-\Omega_{,x}}{(r+r_0)^2}\,\dd r\wedge\dd \tau+\dfrac{\Omega_{,xx}}{r+r_0}\,\dd x\wedge\dd \tau\Big)+\nonumber\\
    &\quad\,\,  \dfrac{\cos\gamma}{2\alpha|c|}\,\Big((r+r_0)\Omega_{,rr}\,\dd r\wedge \dd \phi+\big((r+r_0)\Omega_{,rx}-\Omega_{,x}\big)\,\dd x\wedge\dd \phi\Big),
\end{align}
and find its Hodge dual $\tilde{\mathbf{F}}=*\mathbf{F}$,
\begin{align}\label{dualF}
    \tilde{\mathbf{F}}&=\dfrac{\cos\gamma}{2\alpha |c|}\Big(\dfrac{(r+r_0)\Omega_{,rx}-\Omega_{,x}}{(r+r_0)^2}\,\dd r\wedge\dd \tau-\dfrac{Q}{P}\dfrac{\Omega_{,rr}}{r+r_0}\,\dd x\wedge\dd \tau\Big)+\nonumber\\
    &\quad\,\,\dfrac{\sin\gamma}{2\alpha|c|}\Big(\dfrac{P}{Q}(r+r_0)\Omega_{,xx}\,\dd r\wedge \dd \phi-\big((r+r_0)\Omega_{,rx}-\Omega_{,x}\big)\,\dd x\wedge\dd \phi\Big).
\end{align}

This allows us to calculate the corresponding \emph{electric charge} and \emph{magnetic charge}
\begin{align}
    q_e :=&-\dfrac{1}{4\pi}\int_0^{2\pi C}\!\!\!\int_{-1}^1 \tilde{F}_{x\phi}\,\dd x\,\dd\phi\,,\label{def-qe}\\[4mm]
    q_m :=&-\dfrac{1}{4\pi}\int_0^{2\pi C}\!\!\!\int_{-1}^1 F_{x\phi}\,\dd x\,\dd\phi\,,\label{def-qm}
\end{align}
respectively, where $C$ is the conicity parameter introduced in \eqref{range-of-phi}. Direct integrations give \cite{supp_mat}
\begin{align}
    q_e=-C\,\dfrac{r_0}{2|c|}\sin\gamma\,,\qquad
    q_m=C\,\dfrac{r_0}{2|c|}\cos\gamma\,. \label{charges}
\end{align}
Recall that $\gamma$ is the duality-rotation parameter, mixing the electric and magnetic charges. The above integration was performed on a fixed $\tau$ and $r$ (say, the compact black hole horizon), but because the Maxwell equations are source-free the integrals \eqref{def-qe}, \eqref{def-qm} are \emph{independent} of their specific choice.

Let us also notice an interesting fact: In the static limit ${\omega=0}$ the exact physical charges $q_e$ and $q_m$ are \emph{directly related} to the Pleba\'{n}ski--Demia\'{n}ski parameters $e$ and $g$, introduced previously in Eq.~(111) of \cite{Ovcharenko2025}, by simple formulas
\begin{align}
    q_e=C\,e \,, \qquad  q_m=C\,g\,.
\end{align}
The parameters $e$ and $g$ thus indeed represent the electric and magnetic charges of the black hole without rotation, while for rotating black holes the relation becomes more complicated.

\section{Another GP form of the metric, and its subclasses}
\label{sec:alter}

Another Griffiths--Podolsk\'{y} form of these black holes is obtained by a coordinate shift
\begin{align}\label{shift}
\tilde{r} := r+r_0\,,
\end{align}
applied to the metric \eqref{metric-nontwist-GP-r0}. It gives the line element
\begin{align}
    \dd s^2&=\dfrac{1}{\Omega^2}\Big[-{Q}\,\dd\tau^2
           +\dfrac{\dd \tilde{r}^2}{{Q}}
           +\tilde{r}^2\Big(\dfrac{\dd \theta^2}{{P}}+{P}\sin^2\theta \,\dd\phi^2 \Big)\Big],
           \label{metric-nontwist-GP}
\end{align}
where
\begin{align}
{P}&=1-2\alpha m\cos\theta+4\alpha^2m^2 |c|^2 \cos^2\theta\,,\nonumber\\
{Q}&= \Big(1-\frac{r_0}{\tilde{r}}\Big)
    \Big[ (1-4\alpha^2m^2|c|^2)\,\Big(1-\frac{r_0}{\tilde{r}}\Big) -\frac{2m}{\tilde{r}}\, \Big]
    \big[1-\alpha^2(1-4|c|^2)\,(\tilde{r}-r_0)^2\big],   \label{metric-functions-nontwist-GP}\\
\Omega^2&= \big(1-\alpha (\tilde{r}-r_0)\cos\theta\big)^2
+4\alpha^2 |c|^2 (\tilde{r}-r_0)\big[(\tilde{r}-r_0) {P}-(\tilde{r}-r_0-2m)\cos^2\theta\big],
\nonumber
\end{align}
in which we have defined
\begin{align}\label{def-Q}
Q:=\frac{\mathcal{Q}}{\,\tilde{r}^2}\,,
\end{align}
with the electromagnetic field
\begin{align}
  \mathbf{A}& = \dfrac{1}{4\alpha\,\bar{c}}\, \Big[
     \,\im\,\dfrac{\Omega_{,\theta}}{\tilde{r}\sin\theta}\,\dd \tau +\big(\tilde{r}\,\Omega_{,\tilde{r}}-\Omega\big)\,\dd\phi
    \Big].   \label{A-nontwist-theta}
\end{align}

Recall that there are \emph{two distinct  subclasses}, namely \eqref{r0-two cases_1} and \eqref{r0-two cases_2}, that is
\begin{align}
    \hbox{ case a)\,:}&\qquad  r_0=0\,,\label{r0-two cases-again_1}\\
    \hbox{ case b)\,:}&\qquad  r_0=\dfrac{8m|c|^2}{1-4|c|^2(1-4\alpha^2m^2|c|^2)}\,.
    \label{r0-two cases-again_2}
\end{align}

Although the function $Q(\tilde{r})$ looks a bit more complicated than $\mathcal{Q}(r)$, the metric \eqref{metric-nontwist-GP} is simpler than \eqref{metric-nontwist-GP-r0}. Actually, it is more suitable for obtaining \emph{all} the particular subclasses of such black holes.

From the explicit metric functions \eqref{metric-function-P-nontwist}--\eqref{Omega-nontwist} or, equivalently, from \eqref{metric-functions-nontwist-GP} it can be seen that by taking ${\alpha=0}$ we \emph{completely remove the electromagnetic parameter $|c|$ from the metric} (in this sense, $|c|$ is not a free parameter in general). We thus obtain \emph{only vacuum solutions}. This unfortunate feature can be resolved by a suitable \emph{redefinition} of the electromagnetic parameter, namely by introducing
\begin{align}
    B :=2\alpha|c|\,.  \label{B_def}
\end{align}
Actually, we have applied this idea already in our original paper \cite{Ovcharenko2025}, while considering the non-accelerating limit ${\alpha \to 0}$ with ${|c| \to \infty}$ of the uncharged black holes in Sec.~VIII.~E.~3 by keeping $B$ fixed. By this procedure we obtained the new Kerr--BR class of exact spacetimes representing the Kerr black hole immersed in an external uniform Bertotti--Robinson electromagnetic field \cite{Podolsky2025}.

Nothing prevents us from applying the redefinition \eqref{B_def} in a generic case of (static) \emph{charged black holes} which we are studying here. So, we will replace the three parameters $\alpha$,~$m$,~$|c|$ by a set of \emph{three fully independent} parameters $\alpha$, $m$, $B$. As we will demonstrate below, these can by physically understood as representing the acceleration, mass, and electromagnetic field, respectively. (Recall that ${|c|=0}$ gives just vacuum solutions, so that for ${B=0}$ the electromagnetic field vanishes.)

It turns out that such a re-parametrization strongly depends on whether ${r_0=0}$ or ${r_0\neq 0}$.
In fact, due to \eqref{charges}, the first case represents black holes without charges, while the second case describes charged black holes. We will now investigate these two distinct subclasses, namely the case a) and the case b), independently.

\subsection{Uncharged black holes: case a) given by ${r_0=0}$}
\label{sec_r_0_metr}

In this case ${\tilde{r}=r}$, so that
\begin{align}
    \dd s^2&=\dfrac{1}{\Omega^2}\Big[-{Q}\,\dd\tau^2
           +\dfrac{\dd r^2}{{Q}}
           +r^2\Big(\dfrac{\dd \theta^2}{{P}}+{P}\sin^2\theta \,\dd\phi^2 \Big)\Big].
           \label{metric-nontwist-GP-r0=0}
\end{align}
From the definition \eqref{B_def} we get ${4\alpha^2|c|^2=B^2}$, and substituting this expression into the metric functions \eqref{metric-functions-nontwist-GP} we get
\begin{align}
    {P}&=1-2\alpha m \cos\theta+B^2m^2 \cos^2\theta\,, \nonumber\\ 
    {Q}&=\Big(1-B^2m^2-\dfrac{2m}{r}\Big)\big(1+(B^2-\alpha^2)\,r^2\big)\,,\label{PQOmega-r_0=0}\\
    \Omega^2&=(1-\alpha\,r\cos\theta)^2+B^2\big[\,r^2({P}-\cos^2\theta)
       +2m\,r\cos^2\theta\big]. \nonumber 
\end{align}

The Weyl scalar \eqref{psi_notwist} simplifies considerably to
\begin{align}\label{psi_r0=0}
    \Psi_2=&-\dfrac{m}{r^3}\,\Omega^2 \big( 1-\alpha\,r \cos\theta + B^2m \cos^2\theta\big),
\end{align}
and the regularizing conicity given by \eqref{regularity-nontwist} becomes
\begin{align}\label{regularity-r0=0}
C = \frac{1}{1\mp 2\alpha m + B^2 m^2}\,,
\end{align}
so that there are no cosmic strings (or struts) on \emph{both} axes if, and only if, ${\alpha m = 0}$.

These facts allow us to understand the physical meaning of the class of solutions \eqref{metric-nontwist-GP-r0=0} with  \eqref{PQOmega-r_0=0}: It can be interpreted as an  \emph{accelerating uncharged black hole in the external Bertotti--Robinson magnetic field}. Such an interpretation is based on considering the special subcases, such that the three free parameters ${m, B, \alpha}$ are (separately) set to zero. This directly yields the following:

\subsubsection{No mass: ${m=0}$ gives Bertotti--Robinson}
\label{sec:r=0,m=0}

For ${m=0}$ the metric functions simplify to
\begin{align}
    {P}&=1\,,\nonumber\\
    {Q}&=1+(B^2-\alpha^2)\,r^2\,,\label{BR_2}\\
    \Omega^2&=(1-\alpha\,r\cos\theta)^2+B^2 r^2\sin^2\theta\,,\nonumber
\end{align}
and \eqref{psi_r0=0} gives ${\Psi_2=0}$. Being conformally flat, this is the \emph{Bertotti--Robinson} spacetime with a uniform magnetic (or electric) field whose value is determined by $B$. Indeed, by a coordinate transformation ${r, \theta, t \mapsto R, \Theta, \tau}$ given by
\begin{align}
\dfrac{R^2}{\bar{e}^2}=A^2\,\frac{Q}{\,\Omega^2}-1\,, \qquad
\bar{e}\,\sin \Theta=\frac{r}{\Omega}\sin\theta\,, \qquad
\tau=\frac{1}{A}\,t\,,\label{BR-trans}
\end{align}
where
\begin{align}
A^2= \frac{B^2}{B^2-\alpha^2}\,,\qquad   \bar{e} = \frac{1}{B}  \,, \nonumber
\end{align}
the metric \eqref{metric-nontwist-GP-r0=0}, \eqref{BR_2} is put into the usual form of the Bertotti--Robinson spacetime
\begin{align}
    \dd s^2= -\Big(1+\dfrac{R^2}{\bar{e}^2}\Big)\,\dd\tau^2
           +\Big(1+\dfrac{R^2}{\bar{e}^2}\Big)^{-1}\dd R^2
           +\bar{e}^2\Big(\dd \Theta^2+\sin^2\Theta \,\dd\phi^2 \Big),
           \label{metric-BR}
\end{align}
see Eq.~(7.4) in \cite{GriffithsPodolsky:2009}. Notice that big acceleration ${\alpha^2>B^2}$ (implying ${A^2<0}$) results in non-static region ${Q<0}$ of the spacetime \eqref{metric-nontwist-GP-r0=0} for large values of~$r$ above the acceleration horizon. This metric is thus the Bertotti--Robinson spacetime expressed in uniformly accelerating coordinates.

\subsubsection{No magnetic field: ${B=0}$ gives C-metric}
\label{sec:r=0,B=0}

For ${B=0}$ we directly recover the \emph{C-metric} in spherical-like (bi-polar) coordinates \cite{Griffiths2006_2},
\begin{align}
    {P}&=1-2\alpha m \cos\theta\,,\nonumber\\
    {Q}&=\Big(1-\dfrac{2m}{r}\Big)(1-\alpha^2r^2)\,,\\
    \Omega^2&=(1-\alpha\, r\cos\theta)^2\nonumber\,,
\end{align}
see also Eqs.~(14.6), (14.7) in \cite{GriffithsPodolsky:2009}. There is no magnetic field (it is a vacuum solution), and the black hole accelerates due to the presence of cosmic strings (struts) represented by the topological defects, given by \eqref{regularity-r0=0}, located along the non-regular (semi-)axis.

\subsubsection{No acceleration: ${\alpha=0}$ gives Schwarzschild--BR}
\label{sec:r=0,alpha=0}

For ${\alpha=0}$ we immediately obtain the \emph{Schwarzschild--BR solution} which we found and presented in \cite{Podolsky2025}, see Sec.~II.C therein, namely
\begin{align}
    {P}&=1+B^2m^2 \cos^2\theta\,,\nonumber\\
    {Q}&=\Big(1-B^2m^2-\dfrac{2m}{r}\Big)(1+B^2r^2)\,,\label{Schw_BR_2}\\
    \Omega^2&= 1+B^2\big[\,r^2\sin^2\theta + (2m\,r +B^2m^2r^2 )\cos^2\theta\,\big].\nonumber
\end{align}
It represents non-accelerating black holes in an external magnetic field, with completely regular axes.

\subsubsection{General metric}

Returning to the \emph{general} class of uncharged ${r_0=0}$ solutions \eqref{metric-nontwist-GP-r0=0}--\eqref{PQOmega-r_0=0} with three independent parameters $m, B, \alpha$, we can thus indeed interpret them as  \emph{accelerating black holes in external Bertotti--Robinson electromagnetic field}. They can be understood both as \emph{accelerating Schwarzschild--BR} black holes, or \emph{C-metric in an external electromagnetic field}.

The \emph{black hole horizon} is located at \eqref{horiozons-nontwist}, that is
\begin{align}\label{BH-horiozon}
    r_b = \frac{2m}{1-B^2m^2} \,,
\end{align}
independently of the acceleration parameter~$\alpha$. It is always \emph{bigger} than the Schwarzschild horizon at $2m$, and for the \emph{extremal value} ${B=1/m}$ it actually becomes \emph{infinite}. Assuming ${m>0}$, for ${B>1/m}$ there is no horizon in the region ${r>0}$.

For ${\alpha>B}$ there is another zero in the expression \eqref{PQOmega-r_0=0} for the metric function $Q(r)$. This indicates the presence of \emph{another horizon} at
\begin{align}\label{accel-horiozon}
r_a = \frac{1}{\sqrt{\alpha^2-B^2}}\,,
\end{align}
the \emph{acceleration horizon}. Notice that it is independent of the mass $m$ of the black hole. Also, in such a case ${\alpha>B}$ the transformation \eqref{BR-trans} to the usual form of the Bertotti--Robinson spacetime (with ${m=0}$) is not applicable because ${A^2<0}$.

Interestingly, there also exists a \emph{special subcase} ${\alpha=B}$. Then the black hole is given by the metric functions
\begin{align}
    {P}&= (1-B m \cos\theta)^2\,,\nonumber\\
    {Q}&= 1-B^2m^2-\dfrac{2m}{r}\,,\label{alpha=B}\\
    \Omega^2&= 1- 2 \,(1-Bm\cos\theta)\,B\,r \cos\theta + (1-Bm\cos\theta)^2\,B^2 r^2 .\nonumber
\end{align}
The regularizing conicity \eqref{regularity-nontwist} in this special case becomes
\begin{align}\label{regularity-r0=0}
C =  \frac{1}{(1\mp B m)^2}\,,
\end{align}
where the minus sign applies to the axis ${\theta=0}$, while the plus sign applies to ${\theta=\pi}$. There are thus no cosmic strings (or struts) on both axes if, and only if, ${B = 0}$ or ${m = 0}$. It is consistent with the fact that the acceleration $\alpha$ (here equal to the value of the field~$B$) is caused solely by the conical singularities along the axes --- these black holes are uncharged, and thus there is \emph{no Lorentz force}. There is no direct interaction with the external electromagnetic field that would \emph{also} cause acceleration. This will appear in the next case b).

\subsection{Charged black holes: case b) given by ${r_0\neq 0}$}
\label{sec_r_neq_0_metr}

Recall the expressions \eqref{charges} for the physical \emph{electric charge}~$q_e$ and the \emph{magnetic charge}~$q_m$ of the black hole, namely
\begin{align}
    q_e=-C\,\dfrac{r_0}{2|c|} \sin\gamma\,,\qquad
    q_m= C\,\dfrac{r_0}{2|c|} \cos\gamma\,, \label{charges-again}
\end{align}
where $\gamma$ is the duality-rotation parameter. The \emph{value of the ``electromagnetic'' charge} is thus
\begin{align}
    q :=C\,\dfrac{r_0}{2|c|}\,, \label{def-q}
\end{align}
so that ${\,q_e = -q \,\sin\gamma\,}$ and ${\,q_m = q \,\cos\gamma\,}$, implying
\begin{align}
    q^2 = q_e^2 + q_m^2 = C^2(e^2+g^2) = C^2 \dfrac{r_0^2}{4|c|^2}\,. \label{q^2}
\end{align}
The case b) thus represents the family of \emph{charged black holes} with ${q \not =0}$.

Let us emphasize that in the definition \eqref{def-q} we admit that the charge $q$ can be either positive or negative. This is because  the interaction of the charge of the black hole  with the external electromagnetic field obviously depends on the \emph{relative sign} between them, and it influences the corresponding metric functions (unlike in the case of Reissner--Nordstr\"om black holes for which the sign of $q$ does not change the metric functions). Thus, to take this important \textit{physical} property into account, we allow $q$ to take \emph{any sign}.

It may seem that this issue can be simply resolved by an appropriate choice of the duality rotation parameter $\gamma$ that would change the sign of the electric and magnetic charges $q_e$ and~$q_m$. However, it is not so because such a change of the electric and magnetic charges using~$\gamma$ also changes the external electromagnetic field, rotating it between the purely magnetic and purely electric fields (see \cite{Podolsky2025}). Indeed, by changing ${\gamma\mapsto\gamma+\pi}$ we not only change the sign of the charges of the black hole itself, but we also change the sign of the external field, while their relative sign remains the same. In order to overcome this coupling, we allow $q$ to have any sign.

Moreover, it should be emphasized that for ${\gamma=0}$ the expressions \eqref{F} and \eqref{charges-again} give \emph{purely magnetic} field $\mathbf{F}$ and ${\,q_m = q \,}$ (with ${q_e=0}$), while for ${\gamma=\pi/2}$ we get \emph{purely electric} field $\mathbf{F}$ and ${\,q_e = -q \,}$ (with ${q_m=0}$). It means that either we have a \emph{magnetically charged} (dyonic) black hole in an external \emph{magnetic field}, or en \emph{electrically charged} black hole in an external \emph{electric field}. This is an important note to avoid possible confusion, as usually $q$ describes \textit{purely electric} charge of a black hole, while $B$ represents \textit{purely magnetic} external field.

Actually, the reason for the fact why we did not get such a more physical situation of the electrically charged black hole in the external magnetic field is that while taking the static limit ${\omega \to 0}$, we naturally required the parameter $x_0$ to \emph{remain constant} (see Sec.~\ref{section-no twist}). Because of this, the constraint \eqref{r0x0_cond} became degenerate, fully eliminating~$x_0$ from this condition (and then from all the metric functions). This made an additional duality rotation parameter $\beta$, introduced in \cite{Ovcharenko2025}, redundant. Thus, the usual duality rotation between the electric and magnetic charges of the black hole itself became impossible.
\vspace{2mm}

Our idea now is to express the parameters $|c|$ and $\alpha$ in the spacetime \eqref{metric-functions-nontwist-GP} by the physical parameters $B$ and $q$ representing the values of the \emph{external electromagnetic field} and the \emph{black hole electromagnetic charge}, respectively.

By combining \eqref{def-q} with \eqref{r0-two cases-again_2} we get the following relation for ${|c|}$,
\begin{align}
   \dfrac{2q}{C}\,|c| = r_0=\dfrac{8m|c|^2}{1-4|c|^2(1-4\alpha^2|c|^2m^2)}\,.
\end{align}
Substituting ${B^2 = 4\alpha^2|c|^2}$, cf.~\eqref{B_def}, we thus obtain
\begin{align}
  4Iq\,|c|^2 + 4 Cm\,|c| -q = 0\,,
\end{align}
where
\begin{align}
    I := 1-B^2m^2\,. \label{def-I}
\end{align}
This quadratic equation for $|c|$ has \emph{two possible roots}, namely
\begin{align}
    |c_\pm|= \dfrac{C m}{2Iq} \bigg(\pm \sqrt{1+\frac{I q^2}{(Cm)^2}} -1 \,\bigg).\label{c_abs}
\end{align}
The expression under the square root must be non-negative, so that there is a constraint of the magnitude of the electromagnetic field
\begin{align}
    B^2 m^2 \leq  1 + C^2\,\dfrac{m^2}{q^2}\,.
\end{align}
Otherwise, the global structure of the spacetime would strongly differ from the realistic one.

Moreover, by definition, $|c|$ is also non-negative. In view of \eqref{c_abs}, this fact implies that for the positive electromagnetic charge ${q>0}$ we have to take the \emph{plus sign} in \eqref{c_abs}, while for ${q<0}$ we have to take the \emph{minus sign}. It means that to any value of the charge $q$ (positive or negative) a \emph{unique value}~${|c_+|}$ or ${|c_-|}$ is associated, respectively. Later we will see the physical difference between these two distinct cases.

Now, the relation \eqref{B_def} \emph{uniquely fixes the value of the acceleration} to ${\alpha_\pm:=B/(2|c_\pm|)}$,
\begin{align}\label{alpha+-}
    \alpha_\pm = \dfrac{B I q}{Cm}\, \Big( \pm \sqrt{1+Iq^2/(Cm)^2} -1 \Big)^{-1}\,.
\end{align}
From \eqref{def-q} we then easily obtain
\begin{align}\label{r+-}
    r_{0\pm} = \dfrac{m}{I} \Big( \pm \sqrt{1+Iq^2/(Cm)^2} -1 \Big)\,,
\end{align}
so that, interestingly,
\begin{align}\label{Newton}
    C\,r_{0\pm}\, \alpha_\pm = q\, B\,.
\end{align}
Moreover, the following useful identity holds,
\begin{align}\label{r0-identity}
    I\,r_{0\pm}^2 + 2m\,r_{0\pm} = \frac{q^2}{C^2} = e^2+g^2 \,.
\end{align}

\subsubsection{General metric}\label{sec:gen}

This enables us to rewrite the \emph{general metric} \eqref{metric-nontwist-GP}, \eqref{metric-functions-nontwist-GP}  as
\begin{align}
    \dd s^2&=\dfrac{1}{\Omega^2}\Big[-{Q}\,\dd\tau^2
           +\dfrac{\dd \tilde{r}^2}{{Q}}
           +\tilde{r}^2\Big(\dfrac{\dd \theta^2}{{P}}+{P}\sin^2\theta \,\dd\phi^2 \Big)\Big],
           \label{metric-nontwist-GP-again}
\end{align}
where the metric functions have the form
\begin{align}
    {P}&=1-2\,\alpha_{\pm}m \cos\theta+B^2m^2 \cos^2\theta\,, \nonumber\\
    {Q}&=\Big(I-\dfrac{2M_\pm}{\tilde{r}} + \dfrac{e^2+g^2}{\tilde{r}^2}\Big)
          \Big(1+(B^2-\alpha_{\pm}^2)(\tilde{r}-r_{0\pm})^2\Big),\label{Q-metr_funcs_with_B_q}\\
    \Omega^2&= \big(1-(\alpha_\pm\,\tilde{r}-q\,B/C)\cos\theta\big)^2
           +B^2\big((\tilde{r}-r_{0\pm})^2({P}-\cos^2\theta) + 2m(\tilde{r}-r_{0\pm})\cos^2\theta \big),
            \nonumber
\end{align}
in which the parameter ${\,M_\pm:= m+I\, r_{0\pm}\,}$ has the value
\begin{align}\label{def-M}
 M_\pm = \pm \sqrt{m^2+I\,(e^2+g^2)} \,,\qquad
  I = 1-B^2m^2 \,,\qquad
  e^2+g^2 = \dfrac{q^2}{C^2}\,.
\end{align}
Consequently,
\begin{align}\label{r0-next}
    m = \sqrt{M_\pm^2 -I\,(e^2+g^2)}\,,\qquad
    I\,r_{0\pm} = - m  \pm \sqrt{m^2+I\,(e^2+g^2)} \,.
\end{align}

In the derivation of (\ref{c_abs}) we assumed that there exists a relation between the signs of the charge~$q$ and the $\pm$ sign in (\ref{c_abs}) and (\ref{alpha+-}), required to get ${|c_\pm|\ge0}$ which implies ${\alpha_\pm\ge0}$. However, such a constraint \emph{is not} required  because in the resulting metric (\ref{metric-nontwist-GP-again}), (\ref{Q-metr_funcs_with_B_q}) and in the electromagnetic field there are no  terms containing ${|c_\pm|}$. Both the mass $M_\pm$ and the charge $q$ of the black hole thus can have \emph{any unrelated signs}. Also the acceleration parameter $\alpha_\pm$ can be positive, zero, or negative (but we will typically assume that ${\alpha_\pm\ge0}$).

Let us now discuss the \emph{range of values} of the parameter $M_{\pm}$. From \eqref{def-M} it follows that
\begin{align}
    M_{\pm}=\pm \sqrt{(e^2+g^2)+m^2[1-B^2 (e^2+g^2)]}\,.\label{eq_100}
\end{align}

If ${|B|\sqrt{e^2+g^2}<1}$ we see that  then $M_{\pm}$ has a \emph{lower} bound ${\sqrt{e^2+g^2}}$ (in the case of the $+$ sign) or an \emph{upper} bound ${-\sqrt{e^2+g^2}}$ (in the case of the $-$ sign), respectively. As we will show below in Sec.~\ref{sec:m=0,qnot=0}, these bounds correspond to \emph{extreme black holes}. The existence of such bounds on $M_{\pm}$  prevents us from having naked singularities in the case when ${|B|\sqrt{e^2+g^2}<1}$ because the effective mass $M_{\pm}$ cannot be smaller (for the $+$ sign) or greater (for the $-$ sign) than the charges ${\pm\sqrt{e^2+g^2}}$. However, this restriction can be bypassed in the case ${B=0}$, see Sec.~\ref{sec:B=0}.

On the other hand, if ${|B|\sqrt{e^2+g^2}>1}$  then $M_{\pm}$ is bounded from above (for the $+$ sign) or from below (for the $-$ sign). In this case, the effective mass $|M_{\pm}|$ is always smaller than the charges, and we obtain \emph{only a naked singularity}. Because this case does not describe black holes, from now on we only focus on the case when ${|B|\sqrt{e^2+g^2}<1}$.

Now, due to a simple factorized form of \eqref{Q-metr_funcs_with_B_q} it is easy to determine the position of the \emph{two black-hole horizons} given by ${Q=0}$ in the generic case, as the roots of the quadratic equation ${I\,\tilde{r}^2-2M_\pm\,\tilde{r} + (e^2+g^2)}$, which gives us
\begin{align}\label{BH-horiozons}
   \tilde{r}_{b+}  &= \dfrac{1}{I} \Big( M_\pm + \sqrt{M_\pm^2-I\,(e^2+g^2)} \,\,\Big)\,, \nonumber \\
   \tilde{r}_{b-}  &= \dfrac{1}{I} \Big( M_\pm - \sqrt{M_\pm^2-I\,(e^2+g^2)} \,\,\Big) \,.
\end{align}
(Please notice that the $\pm$ sign in the expressions $\tilde{r}_{b\pm}$ represent the outer and inner \emph{horizons} respectively, while the $\pm$ signs in $M_{\pm}$ represent the \emph{different sign} of the \emph{mass and charge}. We warn about this subtlety to avoid a confusion.) This is in agreement with  \eqref{horiozons-nontwist},
\begin{align}\label{horiozons-nontwist-again}
    r_b = \frac{2m}{I} \,.
\end{align}
Indeed, using \eqref{def-M}, \eqref{r0-next} we get
\begin{align}\label{BH-horiozons-rb}
\tilde{r}_{b+}  & = \dfrac{1}{I} \Big( \pm \sqrt{m^2+I\,(e^2+g^2)} +m \,\Big) =\  r_{0\pm} + r_b \, \nonumber\\
\tilde{r}_{b-}  & = \dfrac{1}{I} \Big( \pm \sqrt{m^2+I\,(e^2+g^2)} -m \,\Big) =\  r_{0\pm} \,,
\end{align}
so that ${\tilde{r}_{b+} = r_b}$ and ${\tilde{r}_{b} = 0}$ for ${r_0=0}$.

A simple combination of  \eqref{BH-horiozons-rb} with \eqref{horiozons-nontwist-again} gives us an interesting relation
\begin{align}\label{rb+_rb-}
\tilde{r}_{b+}  = \tilde{r}_{b-} + \frac{2m}{1-B^2m^2}  \,.
\end{align}

There are also \emph{acceleration horizons} when ${\alpha_{\pm}>B}$, arising from the roots of the second large bracket in \eqref{Q-metr_funcs_with_B_q}, because the black hole is accelerating. The physical cause of this acceleration is the \emph{Lorentz force} due to the electromagnetic interaction between the charge~$q$ of the black hole and the external field~$B$, and also the presence of \emph{cosmic strings} (or struts) with internal tension located along the axes, identified by nontrivial conicity $C$.

In this context, it is necessary to investigate the \emph{regularity of the axes} ${\theta=0}$ and ${\theta=\pi}$. In view of $P$ in \eqref{Q-metr_funcs_with_B_q}, they are \emph{both regular} if and only if ${\alpha_{\pm} m=0}$, and the conicity parameter \eqref{regularity-nontwist} is chosen as
\begin{align}
\label{regularity-general}
C = \frac{1}{1+B^2m^2}\,.
\end{align}
By inspecting the expression \eqref{alpha+-}, the regularity condition ${\alpha_{\pm} m=0}$ can only be satisfied if
\begin{align}
B=0\,, \qquad \hbox{or} \qquad  q=0\,, \qquad \hbox{or} \qquad  m=0 \,.
\end{align}
These conditions lead to three natural subcases, namely \emph{no field} (${B=0}$), \emph{no charge} (${q=0}$), and \emph{no mass} (${m=0}$). They will be studied below.

Let us observe that another apparent possibility for having both regular axes by choosing ${I=0}$ does \emph{not} give ${\alpha_{\pm}=0}$ because the identity \eqref{Newton}, that is ${ \alpha_\pm m = B\,q\,m / (C\,r_{0\pm})}$, implies  ${\alpha_\pm m \ne 0}$ unless ${B\,q\,m=0}$ (see Sec.~\ref{sec:I=0} for more details).

\vspace{2mm}

We will now investigate all these important special subcases separately in the corresponding subsections. This will support our physical interpretation of the general metric \eqref{metric-nontwist-GP-again}, \eqref{Q-metr_funcs_with_B_q} as \emph{accelerating charged (Reissner--Nordstr\"om) black holes in an external Bertotti--Robinson electromagnetic field}.


\subsubsection{No magnetic field: ${B=0}$ gives RN}
\label{sec:B=0}

In such a case \textit{without the magnetic field} (${B=0}$) there is also \emph{no acceleration}. Indeed, \eqref{alpha+-} implies ${\alpha_\pm = 0}$. This is physically understandable because the Lorentz force vanishes. The metric functions \eqref{Q-metr_funcs_with_B_q} reduce considerably to
\begin{align}
    {P}=1\,, \qquad \Omega^2 =1\,,\qquad
    {Q}=1-\dfrac{2M_\pm}{\tilde{r}} + \dfrac{e^2+g^2}{\tilde{r}^2}\,, \label{B=0II}
\end{align}
and $r_{0\pm}$ becomes redundant. Because ${P=1}$, both the poles are regular for ${C=1}$, see also \eqref{regularity-nontwist}, and thus ${e^2+g^2=q^2}$. This in obviously the \emph{Reissner--Nordstr\"om} black hole.

Interestingly, we have obtained this result independently of the sign of $q$. This happens because the external magnetic field is absent, and thus only $q^2$ terms appear in the metric. Notice also that, although the Reissner--Nordstr\"om solution in the form \eqref{B=0II} generally admits three possible structures of horizons (namely: two distinct, one degenerate, and none horizon), in the limit ${B \to 0}$ of our solution \eqref{Q-metr_funcs_with_B_q} the case representing a naked singularity cannot occur because ${ M_\pm^2 = m^2+I(e^2+g^2) > I(e^2+g^2)}$ due to the constraint \eqref{def-M}, and thus ${\tilde{r}_{b+} \ne \tilde{r}_{b-} }$. Actually, this is not a problem because we can use $M_\pm$ as the new and more appropriate mass parameter (instead of the original $m$) with \emph{any} value. However, this cannot be done in the general case.

\subsubsection{No charge: ${q=0}$ gives RN--BR$_0$}
\label{sec:q=0}

Let us now discuss the limit when there is \textit{no charge} (${q=0}$), with all other parameters constant. This limit is not so straightforward because the result depends on the \emph{sign} of~$q$.

First, from \eqref{def-M} it follows that for ${q\to 0}$ we get
\begin{align}\label{def-M-for-q=0}
 M_\pm = \pm |m| \,,\qquad
 e^2+g^2 = 0 \,,\qquad
 I = 1-B^2m^2\,.
\end{align}

If ${q>0}$ then we have to choose the \emph{upper sign} in the expressions in the metric functions \eqref{Q-metr_funcs_with_B_q} because of the requirement of positivity of $|c|$ given by  (\ref{c_abs}), implying ${M_+>0}$. But the parameter $\alpha_+$ given by \eqref{alpha+-} \emph{diverges} in the limit ${q\to 0}$ for any fixed~$m$, and so does also all the metric functions, yielding no reasonable result.

Interestingly, if ${q<0}$ then we have to take the \emph{lower sign}, and in such a case the parameters \emph{remain finite} for ${q\to 0}$,  namely
\begin{align}
    \alpha_- = 0\,, \qquad
    r_{0-}=\dfrac{2M_-}{1-B^2M_-^2} \,, \qquad
    M_- = - m <0 \,.\label{r_0_Schw_BR_1}
\end{align}
The corresponding metric functions \eqref{Q-metr_funcs_with_B_q}  become
\begin{align}
    {P}&= 1+B^2M_-^2 \cos^2\theta\,, \nonumber\\
    {Q}&=\Big(1-B^2M_-^2-\dfrac{2M_-}{\tilde{r}}\Big)\big(1+B^2(\tilde{r}-r_{0-})^2\big),\label{q=0-subcase}\\
        \Omega^2&=1+B^2(\tilde{r}-r_{0-})^2\sin^2\theta +B^2M_-
           \big[B^2M_- (\tilde{r}-r_{0-}) - 2 \big](\tilde{r}-r_{0-})\cos^2\theta.\nonumber
\end{align}

By setting ${B=0}$, one directly obtains the \emph{Schwarzschild black hole}, but with a \emph{negative mass}  ${M_-<0}$.
For ${M_-=0}$ there is necessarily ${r_0=0}$ implying ${\tilde{r}=r}$, and one gets ${{P}=1}$, ${{Q}=1+B^2\,r^2}$, ${\Omega^2=1+B^2\,r^2\sin^2\theta}$. This is the \emph{Bertotti--Robinson spacetime}, see the metric functions~\eqref{BR_2} for ${\alpha=0}$.

The metric \eqref{metric-nontwist-GP-again} with the functions \eqref{q=0-subcase} thus represents an uncharged (${q=0}$) Reissner--Nordstr\"om--Bertotti--Robinson black hole (which we will abbreviate as RN--BR$_0$) with ${M_-<0}$ and ${ r_{0-}=2M_-/(1-B^2M_-^2)\ne0}$. Surprisingly, it is \emph{distinct} from the Schwarzschild--BR black hole with ${r_0=0}$ and ${M=m}$ given by \eqref{Schw_BR_2}.

Both the axes ${\theta=0,\pi}$ are regular for the choice ${C = 1/(1+B^2M_-^2)}$. Then the acceleration parameter also vanishes, ${\alpha_- = 0}$, in agreement with the fact that there is no physical source for acceleration (the cosmic string is absent, and the black hole charge is zero).

\subsubsection{The limit ${m \to 0}$ with ${q \to 0}$ gives Bertotti--Robinson}
\label{sec:m=0,q=0}

Another case we investigate is the limit in which there is \emph{no black hole}, and we are left with just the background metric. We expect to obtain the Bertotti--Robinson spacetime. However, performing the limit ${m \to 0}$ and ${q \to 0}$ is more delicate because we deal with two parameters that are not fully independent and unconstrained.

The case ${q<0}$ is simple. In such a situation it was shown in previous subsection that the limit ${q\to 0}$ leads to the metric functions \eqref{q=0-subcase}.  Due to \eqref{r_0_Schw_BR_1}, the limit ${m \to 0}$ gives ${M_- \to 0}$ and ${ r_{0-} \to 0 }$, so that the expressions \eqref{q=0-subcase} indeed reduce to the metric functions of the Berotti--Robinson spacetime given by \eqref{BR_2} for the case ${\alpha=0}$.

The complementary case ${q>0}$ is a bit more complicated. For any fixed~$m$ the metric functions diverge as ${q\to 0}$. However, if we take the \emph{simultaneous} limit ${q\to0}$ and ${m\to 0}$ such that their \emph{ratio}
\begin{align}\label{def-gamma}
    \gamma := \frac{q}{m}
\end{align}
remains constant, then the expressions remain finite. Indeed, in such a case, using the fact that ${I=1-B^2m^2\to 1}$ and also that the regularizing conicity \eqref{regularity-r0=0} becomes ${C = 1/(1+B^2m^2) \to 1}$, the acceleration parameter $\alpha_+$ approaches the constant
\begin{align}\label{alpha_+-gamma}
    \alpha_+ = \frac{\gamma B}{\sqrt{1+\gamma^2} - 1 }\,.
\end{align}
Moreover, it follows from \eqref{r+-} and \eqref{def-M} that ${r_{0+}\to 0}$ and ${M_+\to 0}$. Thus, by taking this specific limit of the general metric functions \eqref{Q-metr_funcs_with_B_q} we obtain
\begin{align}
    {P}&=1\,, \nonumber\\
    {Q}&= 1+(B^2-\alpha_{+}^2)\,\tilde{r}^2\ , \label{q/m-subcase}\\
    \Omega^2&= (1-\alpha_+\,\tilde{r}\cos\theta)^2 +B^2 \tilde{r}^2 \sin^2\theta\,. \nonumber
\end{align}
This is exactly the metric \eqref{BR_2} of conformally flat Bertotti--Robinson spacetime. In addition to the parameter~$B$ representing \emph{constant value of the (electro)magnetic field}, there is the second parameter $\alpha_+$ representing the \emph{acceleration} of the ``test particle'' located at ${\tilde{r}=0}$.

This is associated with the \emph{acceleration horizon} at ${r_a= 1/ \sqrt{\alpha_+^2-B^2}}$, see \eqref{accel-horiozon}. It occurs if ${\alpha_+^2>B^2}$ which, using the expression \eqref{alpha_+-gamma}, puts a trivial restriction ${\gamma^2>0}$ on the parameter ${\gamma=q/m}$, the fraction of the charge and the mass. The limiting cases are ${\gamma \to 0}$ resulting in ${\alpha_+ \to \infty}$, and ${\gamma \to \infty}$ resulting in ${\alpha_+ \to B}$. For the special case ${q=m}$, that is for ${\gamma=1}$, we get ${\alpha_+ = B/(\sqrt2-1)}$. Recall also \eqref{def-M-for-q=0}, according to which ${M_+ \to |m|}$ in the  no-charge limit ${q\to 0}$.

We may thus conclude that even though the no-charge limit ${q\to 0}$ does \emph{not} give the Schwarzschild--BR spacetime \eqref{Schw_BR_2}, see section~\ref{sec:q=0}, the \emph{simultaneous limit} ${q\to 0}$ and ${m\to 0}$ with the \emph{fixed} fraction ${\gamma = q/m}$ gives the Bertotti--Robinson spacetime. Moreover, the parameter $\gamma$ is uniquely related to the acceleration $\alpha_+$ by the expression \eqref{alpha_+-gamma} which can be inverted to
\begin{align}\label{gamma(alpha_+)}
   \gamma  = \frac{2\,\alpha_+\,B}{\alpha_+^2 -B^2}\,.
\end{align}

\subsubsection{The limit ${m \to 0}$ with $q$ fixed gives extreme Reissner--Nordstr\"om in BR electromagnetic field}
\label{sec:m=0,qnot=0}

Next, we will consider the subcase of the general metric \eqref{metric-nontwist-GP-again}, \eqref{Q-metr_funcs_with_B_q} in which ${m \to 0}$ while the electromagnetic charge~$q$ of the black hole is \emph{kept constant}. Because ${I \to 1}$ and also the regularizing conicity ${C \to 1}$, from \eqref{alpha+-}, \eqref{r+-} and \eqref{def-M} we get
\begin{align}\label{alpha-m=0-r+-m=0}
    \alpha = B  \,, \qquad    r_{0} = q  = M \,,
\end{align}
so that \eqref{Q-metr_funcs_with_B_q} reduce considerably to
\begin{align}
    {P}&=1\,, \nonumber\\
    {Q}&=\Big(1-\dfrac{M}{\tilde{r}}\Big)^2,\label{Q-m=0}\\
    \Omega^2&= 1 -2B(\tilde{r}-M)\cos\theta + B^2(\tilde{r}-M)^2  \,.    \nonumber
\end{align}

From $Q(\tilde{r})$ it is clear that the \emph{black hole horizons coincide} (degenerate) at
\begin{align}\label{BH-horiozons-extreme}
   \tilde{r}_{b+}  &= \tilde{r}_{b-} = M  \,.
\end{align}
Notice that it is consistent with the relation \eqref{rb+_rb-} according to which
${\tilde{r}_{b+}  = \tilde{r}_{b-} \Leftrightarrow m=0}$, and that on this extremal horizon the conformal factor is  ${\,\Omega^2(\tilde{r}_{b\pm})=1}$. This observation in particular shows that the parameter $m$ does \emph{not always} represent a physical mass of the generic spacetime (\ref{Q-metr_funcs_with_B_q}), (\ref{Q-metr_funcs_with_B_q}), but it is rather the parameter describing the \emph{extremality of the black hole} (for ${m=0}$ we get an extremal black hole). More convenient parameter to be used as the mass is $M_{\pm}$ given by (\ref{def-M}).

Interestingly, there is \emph{no acceleration horizon} in this subcase because the magnetic field~$B$ reaches the special value~$\alpha$.

For ${B(=\alpha)=0}$ we get ${\Omega^2= 1}$, and this gives the standard metric of the extreme Reissner--Nordstr\"om black hole. There is neither a cosmic string nor an electromagnetic field, so that there is no physical cause of acceleration.

For ${M(=q)=0}$  we get ${P=1=Q}$ and ${\Omega^2 = 1 -2B \,\tilde{r}\cos\theta + B^2\tilde{r}^2 }$, which can be rewritten as ${\Omega^2 = (1-B\,\tilde{r}\cos\theta)^2 +B^2 \tilde{r}^2 \sin^2\theta}$. This is the metric \eqref{q/m-subcase} of conformally flat Bertotti--Robinson spacetime for the special value of the acceleration parameter ${\alpha=B}$.

These two subcases justify our physical interpretation of the metric  \eqref{metric-nontwist-GP-again} with \eqref{Q-m=0} as describing \emph{accelerating extreme Reissner--Nordstr\"om black hole in the Bertotti--Robinson field}. The axis is \emph{fully regular}, and the acceleration is solely due to the interaction between the charge $q$ (equal to mass~$M$) of the black hole and the external (electro)magnetic field~$B$.

Interestingly, this solution can also be interpreted a particular subcase of the Majumdar--Papapetrou (MP) solution (see, e.g., Sec.~22.5 in~\cite{GriffithsPodolsky:2009})
\begin{align}
    \dd s^2=-\dfrac{1}{U^2}\,\dd t^2+U^2(\dd x^2+\dd y^2+\dd z^2)\,, \label{metric_MP}
\end{align}
where the function $U(x,y,z)$ is
\begin{align}
    U=\epsilon+\sum_{i=1}^N \dfrac{m_i}{\sqrt{(x-x_i)^2+(y-y_i)^2+(z-z_i)^2}}\,.
\end{align}
(We are grateful to Roberto Emparan for making this observation.) Physical interpretation of such a general MP spacetime is that is represents $N$~extremal Reissner--Nordstr\"{o}m black holes of masses ${m_{i}=q_{i}}$ placed at ${(x_{i},~y_{i},~z_{i})}$, with the horizon geometries determined by ${\epsilon=+1,-1}$, or~$0$. Generic MP spacetime is of algebraic type I, but for ${\epsilon=0}$ and ${N=2}$ it becomes of type D. This suggests a possible equivalence between this specific MP spacetime and the extreme Reissner--Nordstr\"om black hole in the Bertotti--Robinson background considered in this subsection. Indeed, after the transformation
\begin{align}
    x&=\dfrac{(\tilde{r}-M)\sin\theta \cos\varphi}{1-2 B (\tilde{r}-M)\cos\theta+B^2(\tilde{r}-M)^2}\,,\nonumber\\
    y&=\dfrac{(\tilde{r}-M)\sin\theta \sin\varphi}{1-2 B (\tilde{r}-M) \cos\theta+B^2(\tilde{r}-M)^2}\,,\\
    z&=(\tilde{r}-M)\dfrac{\cos\theta-B (\tilde{r}-M)}{1-2 B (\tilde{r}-M) \cos\theta+B^2(\tilde{r}-M)^2}\,, \nonumber
\end{align}
for a \emph{very specific} choice of the parameters
\begin{align}
    m_1&= M\,,\quad x_1=y_1=0\,,\quad z_1=0\,, \nonumber\\
    m_2&= \dfrac{1}{B}\,,\quad x_2=y_2=0\,,\quad z_2=-\dfrac{1}{B}\,,
    \label{MP_params}
\end{align}
the metric \eqref{metric_MP} is put into the form \eqref{metric-nontwist-GP-again} with the functions \eqref{Q-m=0}.

By inspecting the parameters (\ref{MP_params}), this spacetime can be interpreted as  created by two black holes. The first has mass $M$ and it is placed at the origin of \eqref{metric_MP} at ${x=y=z=0}$ (in the spherical-like coordinates used in \eqref{metric-nontwist-GP-again} this corresponds to the (degenerate) horizon ${\tilde{r}=M}$), while the second has the specific value of the mass ${m_2=\dfrac{1}{B}=\dfrac{1}{\alpha}}$ and it is placed at ${x=y=0,~z=-\dfrac{1}{B}=-\dfrac{1}{\alpha}}$ (in the spherical-like coordinates this corresponds to ${\tilde{r}\to \infty}$). Thus, our solution can also be interpreted as an extreme Reissner--Nordstr\"om  black hole which is in equilibrium with the field of the ``large external'' black hole located at ${\tilde{r}=\infty}$. The above specific fine-tuning of the parameters of this second black hole creates a uniform external electromagnetic field of the Bertotti--Robinson type for the first black hole, which is placed in the near-horizon region of the second black hole.

In fact, such an interpretation provides a new look at the whole large class of spacetimes presented in \cite{Ovcharenko2025}, which can thus be understood as a \emph{non-extremal} black hole in the field of a \emph{larger extremal} black hole located \emph{very far away}. This, in particular, may appear to be helpful while interpreting the asymptotical structure of this larger class.


\subsubsection{The limit ${\alpha \to 0}$: no acceleration gives RN, RN--BR$_0$, and BR}
\label{sec:alpha=0}

Let us investigate the subcase in which the acceleration parameter~$\alpha_\pm$ vanishes. In view of \eqref{alpha+-}, this may only happen if ${B = 0}$, or ${ q = 0}$, or ${m \to 0}$. These possibilities have been investigated in previous subsections.

Recall that the subcase ${B=0}$ gives the Reissner--Nordstr\"om (RN) black hole (Sec.~\ref{sec:B=0}),
the subcase ${q=0}$ gives the uncharged Reissner--Nordstr\"om--Bertotti--Robinson (RN--BR$_0$) black hole (Sec.~\ref{sec:q=0}), and
the subcase ${m=0}$ gives just the Bertotti--Robinson (BR) universe (Secs.~\ref{sec:m=0,q=0} and \ref{sec:m=0,qnot=0}). Notice that it follows from \eqref{alpha_+-gamma} and \eqref{def-gamma} that ${\alpha_+=0}$ implies ${q=0}$ (unless ${B=0}$).

Recall also that it is not possible to perform the limit ${I \to 0}$ such that ${\alpha \to 0}$ while ${r_0}$ remains finite. This immediately follows from the identity \eqref{Newton} that for any nonzero value of ${q\,B}$ we obtain ${r_{0\pm} \to \infty}$ as ${\alpha_\pm \to 0}$.

We thus conclude that the \emph{only} solutions for which the acceleration parameter~$\alpha$ vanishes are the RN or RN--BR$_0$ black holes, and the BR universe without black holes, discussed in previous subsections.

\subsubsection{The limit ${I \to 0}$: accelerating charged black hole with just one horizon}
\label{sec:I=0}

There exists a unique exact solution with a \emph{finite value} of ${r_{0\pm}}$ corresponding to the limit ${I\to0}$, so that ${M_\pm = m}$ and
\begin{align}\label{I=0}
    B= \dfrac{1}{m}\,.
\end{align}
This is obtained by taking the \emph{upper sign} in \eqref{alpha+-} and \eqref{r+-},\footnote{The lower-sign case gives ${r_{0-}\to \infty}$, and thus diverging metric functions.} corresponding to a positive charge ${q>0}$, with ${q^2=C^2(e^2+g^2)}$. In such a case we get the finite limits
\begin{align}\label{r0-for-I=0}
    r_{0} = \dfrac{e^2+g^2}{2m} \,,  \qquad
    \alpha = \dfrac{2}{\sqrt{e^2+g^2} } \equiv \,2\,\frac{C}{q} \,,
\end{align}
consistent with the relation \eqref{Newton}, that is
\begin{align}\label{identity}
    \alpha\,m\,r_{0} = \frac{q}{C} \,,
\end{align}
where we simplified the notation of the parameters to ${r_0 \equiv r_{0+}}$ and ${\alpha\equiv \alpha_+}$.

Since ${\alpha\, m}$ is nonzero, there is a cosmic string on the axis causing the acceleration, in addition to the Lorentz force given by the interaction of the charge~$q$ of the Reissner--Nordstr\"om black hole with the external (electro)magnetic Bertotti--Robinson field with the special value~${B=1/m}$. The value of the acceleration $\alpha$ is also fixed by \eqref{r0-for-I=0}.

In such a specific case, the metric functions \eqref{Q-metr_funcs_with_B_q} of  \eqref{metric-nontwist-GP-again} become
\begin{align}
    {P}&= 1-2 \alpha \,m \cos\theta + \cos^2\theta\,, \nonumber\\
    {Q}&= -\dfrac{2m}{\tilde{r}^2} \big(\tilde{r} - r_{0} \big)
          \Big[\,1+\Big(\dfrac{1}{\,m^2}-\alpha^2 \Big)(\tilde{r}-r_{0})^2\,\Big],\label{Q-metr_funcs_I=0 with alpha}\\
    \Omega^2&= 1 + 2 (\tilde{r}-r_{0})\Big[\,\dfrac{1}{m}\cos^2\theta - \alpha \cos\theta \Big]  + (\tilde{r}-r_{0})^2\Big[\,\dfrac{1}{m^2} \big(1-2 \alpha\,m  \cos\theta\big) + \alpha^2\cos^2\theta \Big]     \,.\nonumber
\end{align}
From ${Q=0}$ it can now be seen that $r_{0}$ is the \emph{position of a single horizon}. This fully agrees with previous expressions \eqref{BH-horiozons} which for ${I\to0}$ give
\begin{align}\label{BH-horiozons-I=0}
   \tilde{r}_{b+} \to \infty \,, \qquad
   \tilde{r}_{b-}  = \frac{e^2+g^2}{2m}\,.
\end{align}

Due to the relations \eqref{I=0} and \eqref{r0-for-I=0} the metric now contains only \emph{two} independent parameters. It is natural to chose them to be $r_0$ and $\alpha$ because they are directly related to geometry of the spacetime, namely the position of the horizon and the acceleration which is proportional to the ``strength'' of the cosmic string. The remaining physical parameters of mass, charge, and electromagnetic field are then
\begin{align}\label{m/eg/B-for-I=0}
    e^2+g^2 = \dfrac{4}{\alpha^2}\,,\qquad
    m = \dfrac{2}{\alpha^2\,r_{0}} \,,  \qquad
    B = \dfrac{\alpha^2\,r_{0}}{2} \,,
\end{align}
and the metric functions become
\begin{align}
    {P}&= 1-\dfrac{4}{\alpha\,r_{0}} \cos\theta + \cos^2\theta\,, \nonumber\\
    {Q}&= -\dfrac{1}{\tilde{r}^2}\big(\tilde{r} - r_{0} \big)\dfrac{4}{\alpha^2\,r_{0}}
          \Big[\,1+\alpha^2 \Big(\dfrac{\alpha^2\,r_{0}^2}{4}-1\Big)(\tilde{r}-r_{0})^2\,\Big],\label{Q-metr_funcs_I=0 with alpha-and-r_0}\\[2mm]
    \Omega^2&= 1 - 2\alpha \, (\tilde{r}-r_{0})\cos\theta\,
          \Big( 1 - \dfrac{\alpha\,r_{0}}{2}\cos\theta \Big)
           + \alpha^2(\tilde{r}-r_{0})^2 \Big(\dfrac{\alpha\,r_{0}}{2} - \cos\theta \Big)^2  \,.
           \nonumber 
\end{align}

Interestingly, the special subcase ${\alpha\,r_0=2}$, which implies ${\alpha\,m=1}$ and ${B=\alpha}$,
gives
\begin{align}
    {P}&= (1-\cos\theta)^2\,, \nonumber\\
    {Q}&= -\dfrac{2m}{\tilde{r}^2} \,(\tilde{r} - r_{0})\,,\label{Q-metr_funcs_I=0 with alpha m=1}\\
    \Omega^2&= 1 - 2\alpha\, (\tilde{r}-r_{0})\cos\theta
\,( 1 -\cos\theta)+ \alpha^2 (\tilde{r}-r_{0})^2(1-\cos\theta)^2  \,.\nonumber
\end{align}
Although ${B=\alpha}$, this is obviously \emph{different} from the case \eqref{alpha-m=0-r+-m=0} because of a different horizon structure of black holes. While $Q$ given by \eqref{Q-m=0} has an \emph{extreme} (double-degenerate) horizon at~${\tilde{r} = r_{0} = M = q }$, the black hole with \eqref{Q-metr_funcs_I=0 with alpha m=1} has just a \emph{single} horizon at ${\tilde{r} = r_{0} = q^2/(2C^2m) }$.

\vspace{4mm}

To conclude, the type~D exact solutions of the Einstein--Maxwell equations  given by the metric \eqref{metric-nontwist-GP-again}, \eqref{Q-metr_funcs_with_B_q} can be physically understood as representing \emph{charged black holes without rotation, accelerating in the Bertotti--Robinson universe}.

Interestingly, in the uncharged situation there are thus \emph{two distinct} spacetimes which are ``Schwarzschild-like'' black holes immersed in the external Bertotti--Robinson electromagnetic field. Indeed, the corresponding class of spacetimes with ${r_0\not=0}$ \emph{is not the same  as the Schwarzschild--BR black hole} arising for ${r_0=0}$, given by \eqref{metric-nontwist-GP-r0=0} with \eqref{Schw_BR_2}, found and presented in Eq.~(18) of~\cite{Podolsky2025}. This may seem strange at first glance, but the physical reason for such a dichotomy will become clear in the next Sec.~\ref{section-VdBC} where we will present an explicit relation to the Van den Bergh--Carminati form of these two \emph{distinct} classes of non-twisting black holes \cite{VandenBergh2020}, namely ${hj = 0}$ and ${hj \neq 0}$, equivalent to our case a)  and case b), that is ${r_0=0}$ and  ${r_0 \neq 0}$.

Finally, let us recall the \emph{exact} relation \eqref{Newton}, namely
\begin{align}
     m_{\rm eff}\,\alpha = q\,B\,,
\end{align}
where we have introduced the constant ${ m_{\rm eff} \equiv C\,r_{0}}$. It \emph{resembles Newton's law} of motion of the black hole of an ``effective inertial mass''~$m_{\rm eff}$ accelerating in the external electromagnetic field, if $\alpha$ denotes the  acceleration, and ${q\,B}$ represents the ``electromagnetic force'' given by the product of the charge and the value of the field.


\section{Relation to the Van den Bergh--Carminati solutions}
\label{section-VdBC}

The above metric forms of the non-twisting (${\omega=0}$) type~D solutions to the Einstein--Maxwell equations with a non-aligned electromagnetic field  are very useful for their physical interpretation, and possibly for other studies. We obtained them here as a special limiting subcase of the more general and larger class \eqref{ds_alpha_omega}--\eqref{A-3} with a general twist~$\omega$, presented in~\cite{Ovcharenko2025}. Actually, our original derivation therein was inspired by the work \cite{VandenBergh2020} of Van den Bergh and Carminati who found \emph{all non-twisting} and non-aligned  Einstein--Maxwell solutions of type~D within the Robinson--Trautman class. It means that our class of solutions \eqref{metric-nontwist-PD}--\eqref{Q-nontwist-PD}, equivalent to \eqref{metric-nontwist-GP-r0}--\eqref{Omega-nontwist}, should be equivalent to the Van den Bergh--Carminati representation, previously found in \cite{VandenBergh2020}.

It is indeed so, as we will now prove by presenting explicit relations between these two forms of the complete class of such spacetimes. Actually, these relations depend on whether ${r_0=0}$ or ${r_0\not=0}$.

\subsection{The case a) given by ${r_0=0}$}

In this simpler case, our general metric \eqref{metric-nontwist-PD}--\eqref{Q-nontwist-PD} reduces to
\begin{align}
    \dd s^2=\dfrac{r^2}{\Omega^2}\Big(-\dfrac{\mathcal{Q}}{r^4}\,\dd\tau^2
           +\dfrac{\dd r^2}{\mathcal{Q}}
           +\dfrac{\dd x^2}{\mathcal{P}}+\mathcal{P}\,\dd\phi^2\Big).
\end{align}
The transformation
\begin{align}
    r=\dfrac{1}{q_1 \,q+q_0}\,,\qquad x=p+p_0\,,\qquad \tau=\dfrac{\eta}{2m}\,,\qquad\phi=\dfrac{q_1}{2m}\, \sigma\,,
\end{align}
where
\begin{align}
    q_1=\dfrac{1}{8\alpha^2 m|c|^2}\,,\qquad
    q_0=\dfrac{\epsilon}{6m}\,,\qquad
    p_0=\alpha\,q_1\,,\qquad
    K^2=\dfrac{q_1}{m}\,,
\end{align}
brings it to the form
\begin{align} \label{VandenBergh--metric-e=1}
    \dd s^2=\dfrac{K^2}{2N}\,\big(\!-\xi^2 \dd \eta^2+\xi^{-2} \dd q^2
           +\varsigma^{-2}\dd p^2+\varsigma^2 \dd \sigma^2\big)\,,
\end{align}
where $\xi^2(q)$ is a quartic function of $q$, $\varsigma^2(p)$ is a quartic function of $p$,  while $N(p,q)$ is quadratic in both  $p$ and $q$.

This metric is exactly the Van den Bergh--Carminati metric Eq.~(85) of \cite{VandenBergh2020} (after renaming the key coordinates therein as ${s \mapsto p}$, ${x \mapsto q}$), where
\begin{align}
    N&=\Big(q-6\dfrac{\Sigma_0}{\Xi_0}\Big)p^2+3\dfrac{Q_0}{\Xi_0}\,p-\dfrac{\Xi_0}{3}\,q^2-2\Sigma_0\, q
        +P_0 -12 \dfrac{\Sigma_0^2}{\Xi_0}\,,\nonumber\\[2mm]
    \xi^2&=\dfrac{\Xi_0}{3}\,q^3-P_0\,q+\dfrac{3}{4\Xi_0^2}(8\Xi_0\Sigma_0 P_0-96 \Sigma_0^3+3Q_0^2)\,,\\[2mm]
    \varsigma^2&=-\dfrac{1}{4}\,p^4+3\Sigma_0 \,p^2-Q_0\,p-\dfrac{\Xi_0}{3}P_0+3\Sigma_0^2\,.\nonumber
\end{align}

We do not provide explicit complicated expressions for the parameters $P_0, Q_0, \Sigma_0, \Xi_0$ in terms of the PD parameters appearing in \eqref{P-nontwist-PD}--\eqref{Q-nontwist-PD} because they are not important (they are just \emph{ad hoc} constants of integration without a straightforward physical relevance). Their explicit expressions in terms of the PD parameters can be found in the supplementary Wolfram file \cite{supp_mat}. It suffices to observe that the conformal factor $N$ is quadratic both in $p$ and~$q$, $\varsigma^2$ is a quartic polynomial of~$p$, while $\xi^2$ is a \emph{cubic} polynomial of~$q$ (this is a consequence of missing constant term in ${\mathcal{Q}}$ given by (\ref{Q-nontwist-PD})). This fully agrees\footnote{We changed the constants $p, q$ of \cite{VandenBergh2020} to $P_0, Q_0$, not to be confused with the coordinates used in \eqref{VandenBergh--metric-e=1}.} with the expressions given by Eqs.~(101)--(103) in \cite{VandenBergh2020},  representing the general \emph{solution in the case} ${h=0}$. Concerning the complementary case ${j=0}$ identified in \cite{VandenBergh2020}, it was noted already by Van den Bergh and Carminati that this is analogous to the case ${h=0}$ due to the transformation given by Eq.~(44) in \cite{VandenBergh2020}, yielding Eqs.~(104)--(106) therein.

Let us finally remark that the complementary case ${e=-1}$ considered in \cite{VandenBergh2020} can be obtained by a trivial redefinition ${\Xi_0\rightarrow -\Xi_0}$.

\subsection{The case b) given by ${r_0\neq 0}$}
To show the equivalence of the solution found in \cite{VandenBergh2020} to our solution in the more complicated case ${r_0\neq 0}$, we start again with the metric (\ref{metric-nontwist-PD})--(\ref{Q-nontwist-PD}). By applying the coordinate transformation
\begin{align}
       r+r_0= \dfrac{1}{q_1\,q+q_0}\,,\qquad
       x= p_1\,p+p_0\,,\qquad
    \tau= \dfrac{\eta}{2m}\,,\qquad
    \phi= \dfrac{q_1}{p_1}\,\dfrac{\sigma}{2m}\,,
\end{align}

\noindent
with the specifically chosen constants
\begin{align}
q_1=&-\dfrac{1}{2}\Bigg(\dfrac{(1-4|c|^2\epsilon)^4}
    {4|c|^2m\big[1+4|c|^2(8\alpha |c|^2m(n'-2 \alpha m |c|^2 k)-\epsilon)\big]^2}\Bigg)^{1/3},\nonumber\\
q_0=&\dfrac{1-4|c|^2\epsilon}
    {32|c|^2m\big[1+4|c|^2(8\alpha |c|^2m(n'-2 \alpha m |c|^2 k)-\epsilon)\big]^2}
    \nonumber\\
    &\hspace{0mm} \times \Big(1-4|c|^2\epsilon-(4|c|^2)^3(4\alpha^2km^2+12 \alpha m(n'-2 \alpha m |c|^2 k)\epsilon-\epsilon^3)+\\
    &\hspace{6mm} +(4|c|^2)^2(12\alpha m(n'-2 \alpha m |c|^2 k)-\epsilon^2)+4 \alpha^2m^2(4|c|^2)^4(4(n'-2 \alpha m |c|^2 k)^2+k\epsilon) \Big),\nonumber\\
p_1&=-\Bigg(\dfrac{1+4|c|^2(8\alpha |c|^2m(n'-2 \alpha m |c|^2 k)-\epsilon)}
    {4\alpha|c|^2m(1-4|c|^2\epsilon)^2}\Bigg)^{1/3},\qquad
p_0=\dfrac{1}{8\alpha |c|^2m}\,,\qquad
K^2=-\dfrac{q_1}{m}\,,\nonumber
\end{align}
we again get the metric  (\ref{VandenBergh--metric-e=1}), but now with
\begin{align}
          N&= q\,p^2-p\,q^2-\dfrac{P_0}{2}(q-p)^2+2\Xi_0(p-2q)+2\Sigma_0(q-2p)+Q_0\,,\nonumber\\
      \xi^2&= \dfrac{1}{4}\,q^4+3\Xi_0\,q^2-[P_0(\Sigma_0+\Xi_0)+Q_0]\,q
          +(2\Sigma_0-\Xi_0)^2+\dfrac{1}{2}P_0Q_0\,,\\
\varsigma^2&=-\dfrac{1}{4}\,p^4+3\Sigma_0\,p^2+[P_0(\Sigma_0+\Xi_0)-Q_0]\,p
          -(2\Xi_0-\Sigma_0)^2-\dfrac{1}{2}P_0Q_0\,.\nonumber
\end{align}
These are exactly the expressions given by  Eqs.~(82)--(83) in \cite{VandenBergh2020} \emph{for the generic case} ${hj\neq 0}$, in which $\xi^2$ is quartic in $q$, $\varsigma^2$ is quartic in $p$, and $N$ is quadratic in both  $p$ and $q$. (Again, we do not provide here explicit expressions for the parameters $P_0, Q_0, \Sigma_0, \Xi_0$ in terms of the PD parameters, but they can be found in the supplementary Wolfram file \cite{supp_mat}).

To summarize, we proved that our solution with nonzero $|c|$ in the non-twisting limit  \eqref{metric-nontwist-GP-r0} --- which is equivalent to  \eqref{metric-nontwist-GP} --- gives all the solutions found in \cite{VandenBergh2020}. Moreover, these metric forms allow us to give a \emph{physical interpretation} to the cases ${hj\neq 0}$, ${h=0}$, and ${j=0}$, identified in the previous systematic work \cite{VandenBergh2020}. As we demonstrated by the above explicit transformations, the case ${hj\neq 0}$ corresponds to our case ${r_0\neq 0}$, where $r_0$ \emph{is the parameter related to the charges $q_e$~and~$q_m$ found by integrating the corresponding fluxes through the horizon surface} \eqref{charges}. On the other hand, the case ${h=0}$ (and also the case ${j=0}$) corresponds to ${r_0=0}$, which  is \emph{equivalent to vanishing of the charges $q_e$ and $q_m$} of the black hole.

Moreover, this one-to-one relation helps us to understand the deeper physical difference between these two subclasses. To discover it, we explicitly express the important GHP functions~$h$ and~$j$ which were introduced above Eq.~(31) in \cite{VandenBergh2020}, namely
\begin{align} \label{def-h-j}
  h \equiv w_0-2g\,,\qquad\quad
  j \equiv f\,\Psi_2+2g-w_0\,,
\end{align}
where
\begin{align} \label{def-h-j}
  g  = \frac{1}{C_0}\,\Phi_1\,,  \qquad
  f  =-\frac{1}{\mu\pi\,C_0}\,\Phi_2\,,\qquad
  w_0= \frac{1}{\mu\,C_0}\,\crpartial \Phi_2\,,
\end{align}
see, respectively, Eqs.~(18), (14), and the combination of expressions below Eqs.~(20), (10) and above Eq.~(1) in \cite{VandenBergh2020}. It gives us
\begin{align}
    h&=\dfrac{\crpartial \Phi_2-2\mu\, \Phi_1}{\mu\,C_0}\,,\label{h_expr}\\[2mm]
    j&=\dfrac{2\mu \pi\, \Phi_1-\pi\,\crpartial \Phi_2-\Phi_2 \Psi_2 }{\mu\pi\,C_0}\,, \label{j_expr}
\end{align}
where $\mu$ and $\pi$ are the optical scalars, ${|C_0|=1}$, $\Phi_A$ are the Maxwell scalars, and $\Psi_2$ is the Weyl scalar. Recall that $\Phi_1$~is the \emph{aligned} component of the electromagnetic field, while ${\Phi_2=\Phi_0}$~is the \emph{non-aligned} component.

It is now obvious that in the case ${h=0}$ there is
\begin{align}
    \Phi_1 = \dfrac{1}{2 \mu}\,\crpartial \Phi_2\,,\label{h=0_expr}
\end{align}
and in the case ${j=0}$ there is
\begin{align}
    \Phi_1 = \dfrac{1}{2 \mu} \Big( \crpartial \Phi_2 +  \dfrac{1}{\pi} \,\Phi_2 \Psi_2\Big). \label{j=0_expr}
\end{align}
Therefore, in the spacetimes with ${hj = 0 \Leftrightarrow r_0 = 0}$ the aligned component~$\Phi_1$ is \emph{determined} by the non-aligned component $\Phi_2$ (and it's derivatives). On the other hand, in the spacetimes with ${hj \ne 0 \Leftrightarrow r_0 \ne0}$ the aligned component $\Phi_1$ and the non-aligned component $\Phi_2$ \emph{are mutually independent} because $\Phi_1$ is \emph{not} simply given by \eqref{h=0_expr} nor \eqref{j=0_expr}. Instead, the relations \eqref{h_expr} and \eqref{j_expr} for both ${h \ne 0}$ and ${j \ne 0}$ keep $\Phi_1$ and $\Phi_2$ unrelated. Notice, in particular, that for the \emph{purely aligned case} ${\Phi_2=0}$ there is ${j=-h= \dfrac{2}{C_0} \Phi_1 \ne 0}$.

To conclude, this provides us with the deeper geometric insight and physical interpretation to the two \emph{distinct classes} of solutions, namely ${r_0=0}$ and ${r_0\neq 0}$. In the case  ${r_0=0}$ of \emph{uncharged} black holes, the aligned component~$\Phi_1$ of the electromagnetic field \emph{is not independent} and can be completely expressed using the non-aligned component~$\Phi_2$ representing the external electromagnetic field. On the other hand, in the complementary case ${r_0\neq 0}$ corresponding to \emph{charged} black holes the aligned component~$\Phi_1$ \emph{is independent} and cannot be determined directly from~$\Phi_2$ by using \eqref{h=0_expr} or \eqref{j=0_expr}.

Such an interpretation is further supported by recalling that the ${r_0=0}$ case represents the \emph{uncharged} black hole in the external electromagnetic field, and in this case both the components $\Phi_2$ and $\Phi_1$ are given by a \emph{single parameter}~$B$ which is the strength of the external field. On the other hand, the ${r_0\neq 0}$ case represents the \emph{charged} black hole in the external electromagnetic field, and there are \emph{two independent parameters}, namely the charge $q$ and the strength of the external field $B$. When ${B\to 0}$, the non-aligned component~$\Phi_2$ of the field completely vanishes (yielding ${h+j=0}$), and only the \emph{independent aligned} component~$\Phi_1$ remains ---  representing the field of just the charged black hole.

Actually, this functional (in)dependence of the aligned and non-aligned components $\Phi_1$ and $\Phi_2$ of the electromagnetic field is the deeper reason why in the \emph{no charge limit} ${q\to0}$ (so that the flux of both the electric and magnetic fields through the black hole horizon becomes zero) we obtain \emph{different classes} of uncharged black holes, namely the
\emph{Schwarzschild--BR spacetime} (with ${r_0 = 0, q=0, M \ne 0}$) which is different from the  \emph{${RN-BR_0}$ spacetime} (with ${r_0 \ne 0, q=0, M<0}$).

\section{Relation to the Alekseev--Garcia and Alekseev solutions}
\label{section-Alexeev}

Finally, we wish to discuss the possible relation of our solutions presented here to the Alekseev--Garcia \cite{Alekseev1996} and the Alekseev solutions \cite{Alekseev2025}.

As we have already mentioned in previous section, our studies \cite{Ovcharenko2025, Podolsky2025} were initially motivated by the work \cite{VandenBergh2020} of Van den Bergh and Carminati from 2020. At that time we were not aware of the 1996 work \cite{Alekseev1996} by Alekseev and Garcia entitled ``Schwarzschild black hole immersed in a homogeneous electromagnetic field''. We were informed about it only after completing and submitting \cite{Ovcharenko2025, Podolsky2025}. Because it was claimed in \cite{Ortaggio2018} that the Alekseev--Garcia spacetime is of algebraic type I, we did not investigate the relation to our class of spacetimes which are, by construction, all of type D.

However, recently the authors of \cite{Ortaggio2018} reconsidered their analysis, and found that the Alekseev--Garcia black hole solution \emph{is actually of type D} \cite{Ortaggio2025}. So it is now important to clarify the relation of the solution \cite{Alekseev2025} to the family of spacetimes we are presenting here.

Unfortunately, this is not easy to do. The reason is that these two classes of metrics were derived by completely different techniques, and have different coordinate representations with different free parameters. More specifically, in \cite{Ovcharenko2025} we have employed the Newman--Penrose formalism, whereas the Alekseev--Garcia 1996 solution was obtained by employing the generating technique developed previously by Alekseev (see \cite{Alekseev1985,Alekseev1988}, summarized in Appendix of \cite{Alekseev1996}). Such a method transforms the problem of finding the Ernst potentials for axially-symmetric electrovacuum spacetimes into the task of solving a specific integral equation in the complex plane. In addition, such an equation acquires two arbitrary holomorphic functions ${\bf u}$ and ${\bf v}$, called the ``monodromy data functions''. Even though this equation is integrable, the analytical solution does not exist for a generic ${\bf u}$ and ${\bf v}$. However, if the complex monodromy data functions ${\bf u}$ and ${\bf v}$ are given in the form of a \emph{ratio of polynomials}, then the analytical solution can be found, in principle.

This interesting generating technique did not attract much attention because it is rather complicated and it is not easy to \emph{explicitly} derive new spacetimes, as it involves non-trivial intermediate steps. However, it already proved to be powerful to generate various exact colliding plane wave spacetimes \cite{Alekseev2000,Alekseev2001,Alekseev2004}. Concerning the black hole spacetimes, this method was applied to generate the Schwarzschild black hole in homogeneous Bertotti--Robinson-type electromagnetic field \cite{Alekseev1996}. Recently it was employed to add a charge to this solution \cite{Alekseev2025}.

Due to the recent correction \cite{Ortaggio2025} of the claim presented in \cite{Ortaggio2018}, these Schwarzschild and Reissner--Nordstr\"{o}m black holes in external electromagnetic field \cite{Alekseev1996, Alekseev2025} are of algebraic type~D. It is thus \emph{likely} that they are equivalent to our solutions studied in this work, or at least that there is a significant overlap between them. However, at present it is not clear how to prove this conjecture. Some suitable invariant classification method could be employed, but it would be the best to find an explicit transformation between the different coordinate systems and physical parameters.

\newpage
Specifically, the Alekseev--Garcia solution was presented in the Weyl-type form \cite{Alekseev1996, Ortaggio2018}
\begin{align}
 \dd s^2=&-e^{2\psi}\cosh^2\frac{z}{b}\,\dd t^2
 +e^{2\gamma}(\dd z^2+\dd\rho^2)+e^{-2\psi}\,b^2\sin^2\frac{\rho}{b}\,\dd\phi^2 ,
 \label{AG}
\end{align}
where the complicated metric functions are
\begin{align}
e^{2\psi}=&\frac{\left(R_+ +R_- -2m\cos\dfrac{\rho}{b}\right)^2}{(R_+ + R_-)^2-4m^2}\,,\nonumber\\[2mm]
e^{2\gamma}=&\frac{\left(R_+ +R_- -2m\cos\dfrac{\rho}{b}\right)^2}{4R_+ R_-}
   \left[\frac{R_+ -b\sinh\dfrac{z}{b}+(l+m)\cos\dfrac{\rho}{b}}{R_- -b\sinh\dfrac{z}{b}+(l-m)\cos\dfrac{\rho}{b}}\right]^2 , \label{AG-coeffs} \\[2mm]
 R_{\pm}^2=&\left(l\pm m-b\sinh\dfrac{z}{b}\cos\dfrac{\rho}{b}\right)^2+b^2\cosh^2\dfrac{z}{b}\sin^2\dfrac{\rho}{b} ,
 \nonumber
\end{align}
in which $m,b,l$ are constants. The electromagnetic field is defined (up to a constant duality rotation $\gamma$) by the potential
\begin{align}
 {\bf A} = -b\,\frac{R_+ +R_- +2m}{R_+ +R_- -2m\cos\dfrac{\rho}{b}}\left(1-\cos\dfrac{\rho}{b}\right)\dd\phi \,.
 \label{A}
\end{align}

It was argued in \cite{Alekseev1996} that the parameter $1/b$ gives the value of the (electro)magnetic field, $m$~(possibly combined with $b$) should correspond to the mass of the black hole (for ${m=0}$ the Bertotti--Robinson universe is obtained), while $l$ determines the location of the black hole along the symmetry axis ${\rho=0}$ of the Weyl coordinates, and also its acceleration due to the conical (nodal) singularities (which are absent for ${l=0}$).

Let us compare \eqref{AG} to our ${r_0 = 0}$ metric \eqref{metric-nontwist-GP-r0=0} for the uncharged black hole accelerating in the  Bertotti--Robinson field, that is
\begin{align}
    \dd s^2&=\dfrac{1}{\Omega^2}\Big[-{Q}\,\dd\tau^2
           +\dfrac{\dd r^2}{{Q}}
           +r^2\Big(\dfrac{\dd \theta^2}{{P}}+{P}\sin^2\theta \,\dd\phi^2 \Big)\Big],
           \label{accelerating uncharged BH in BR}
\end{align}
with the metric functions $P, Q, \Omega $ given by \eqref{PQOmega-r_0=0} depending on \emph{three physical parameters}, namely  $B, m, \alpha$. It is most natural to expect that $B$ should correspond to $1/b$, $m$ could be the same, and $\alpha$ should be related to $l$. However, the exact relation is not clear. Even in the simpler case of the Schwarzschild--BR black hole \eqref{Schw_BR_2} without acceleration ($\alpha=0$ corresponding to ${l=0}$), this looks like a formidable task.

This is further complicated by another principal problem. Namely, it is \emph{not clear at all} that the Alekseev--Garcia solution \eqref{AG-coeffs} is related to our Schwarzschild--BR metric \eqref{accelerating uncharged BH in BR} with ${r_0=0}$. It can also be related to the different RN--BR$_0$ vacuum solution \eqref{metric-nontwist-GP-again} with \eqref{q=0-subcase} that represents uncharged (${q=0}$) Reissner--Nordstr\"om--Bertotti--Robinson black hole with ${M_-<0}$ and ${ r_{0-}=2M_-/(1-B^2M_-^2)}$. The existence of \emph{two different} algebro-geometric structures of these black hole spacetimes, distinguished by whether ${r_0=0}$ or ${r_0\neq 0}$, demonstrated in the current work (in particular at the end of previous Sec.~\ref{section-VdBC}), indicate that there may exist richer families of exact solutions than those explicitly presented in \cite{Alekseev1996} and \cite{Alekseev2025}.

In this context, let us also remark that the exact solutions discussed in our work are different from the ones presented in \cite{Halilsoy1993}. The reason is that the spacetime \cite{Halilsoy1993} has an \emph{aligned} electromagnetic field, and thus it is actually a specific coordinate system for the Reissner--Nordstr\"{o}m black hole. On the other hand, the RN--BR solution discussed in our work here is a \emph{generalization} of the RN spacetime obtained by adding an external \emph{non-aligned} Bertotti--Robinson electromagnetic field.

After the non-trivial issues mentioned above are resolved, relations between the more general families of \emph{charged} black holes in the external BR electromagnetic field can be studied. We postpone these rigorous investigations to a future work.

\section{Summary of the results, and conclusions}
\label{sec:final}

\begin{figure}
    \centering
   \includegraphics[width=0.85\linewidth]{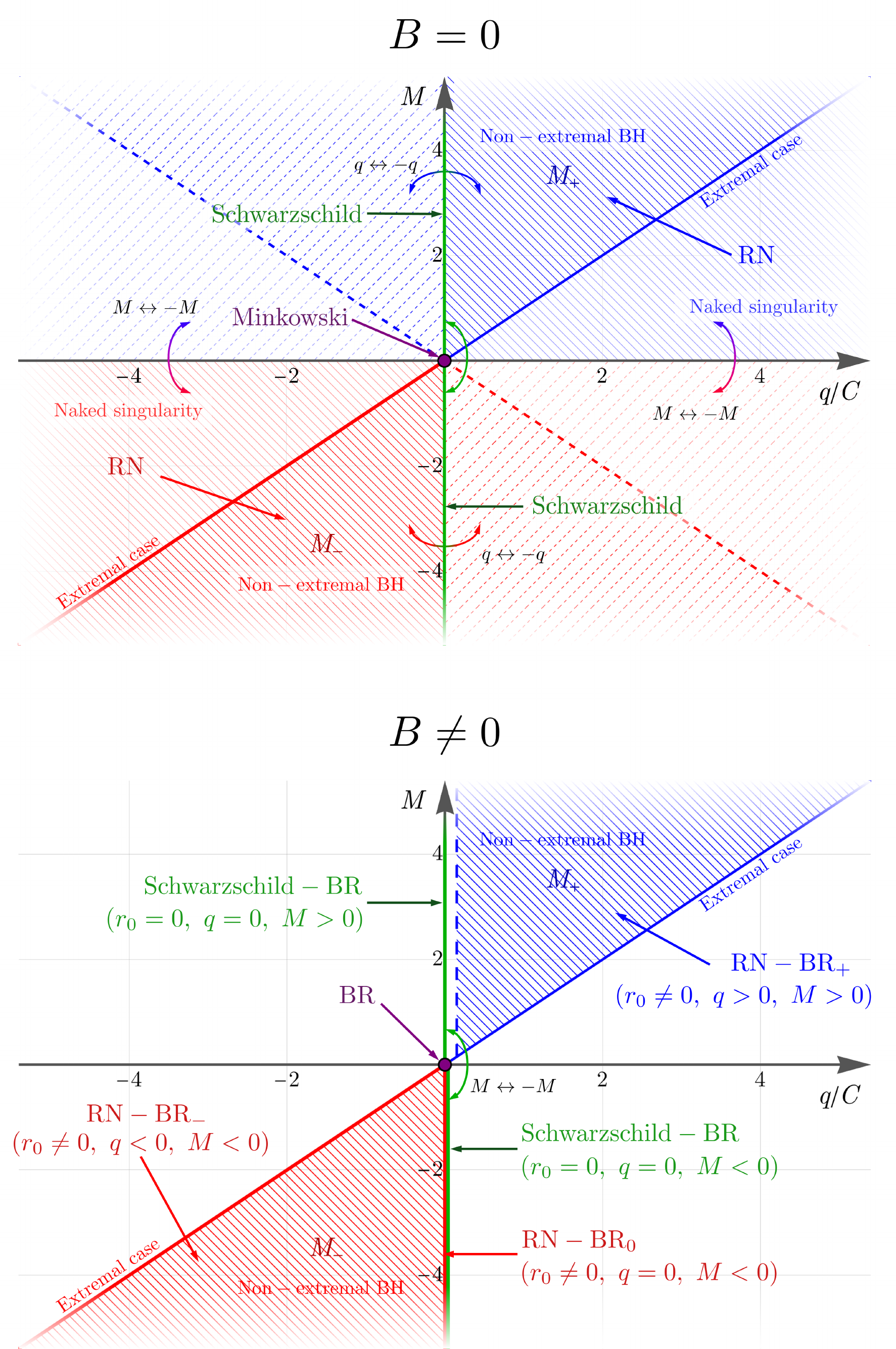}
    \caption{Schemes summarizing all possible charged and accelerating black holes, and their special cases,
     in the external non-aligned electromagnetic field (${B\neq 0}$, the lower part) and without the field (${B=0}$, the upper part), which we derived and described in this paper.}
    \label{FIG_scheme}
\end{figure}

After we have described and analyzed all the special subcases of the non-twisting  black hole spacetimes within the large class of type~D solutions with a non-aligned electromagnetic field, as derived in \cite{Ovcharenko2025}, showing also their explicit relation to the equivalent solutions found previously in \cite{VandenBergh2020}, we now summarize them and present a full picture of these possible subcases. The corresponding complete scheme, containing the main information about all the subcases for different possible masses~$M$ of the black holes (plotted vertically) and their charges $q$ (plotted horizontally) is given in Fig.~\ref{FIG_scheme}. It is divided onto two parts: the upper one represents the case without the external electromagnetic field (${B=0}$), while the lower one summarizes the general case with such a field (${B\neq 0}$).

We start with the description of the lower part of the scheme, representing \emph{charged  black holes accelerating in the external electromagnetic field} ${B\neq 0}$. It contains \emph{blue} and \emph{red triangular domains} that correspond to a generic Reissner--Nordstr\"{o}m black hole immersed in the external Bertotti--Robinson spacetime, abbreviated as RN--BR, presented in Sec.~\ref{sec:gen}. As was discussed in detail there, depending on the \emph{sign of the charge}~$q$ one gets two different types of solutions: the first with a positive mass ${M_+>0}$ if ${q>0}$, and the second with a negative mass ${M_-<0}$ if ${q<0}$, provided ${\alpha_\pm \ge0}$ (see the discussion below Eq.~(\ref{r0-next})). To distinguish these two distinct regions in the parameter space, we denote them as RN--BR$_+$ for the positive charge, and RN--BR$_-$ for the negative charge. The fact that the sign of the charge~$q$  also uniquely defines the sign of the mass parameter $M_{\pm}$ is reflected in the fact that the corresponding RN--BR$_\pm$ are present only in the corresponding quarters with both positive/negative charge and mass. Moreover, these subcases cannot be obtained one from the other by simply inverting the charge (unlike in the case \emph{without} the external field, as we will discuss below).

Another important property reflected in this picture is that if ${|B|\sqrt{e^2+g^2}<1}$ then the mass parameter $M_{+}$ \emph{is bounded from below} (and, analogously, $M_-$ is bounded from above), see the discussion of Eq.~\eqref{eq_100}. Such a limiting value (corresponding to ${m=0}$) gives the \emph{extremal case} studied in detail in Sec.~\ref{sec:m=0,qnot=0}. These extremally charged RN--BR black holes (with ${q=M}$ and special acceleration ${\alpha=B}$) correspond to the thick inclined blue and red lines, limiting the triangular $M_{+}$ and $M_-$ regions.

Another important subcases that are represented in our scheme are the corresponding black holes \emph{without charges}, that is  ${q=0}$. As was discussed in Sec.~\ref{sec:q=0}, the limit ${q\to 0}$ is not well defined for the RN--BR$_+$ subcase (this we indicate by the vertical \emph{dashed} blue line, meaning that such a limit cannot be taken). However, for the RN--BR$_-$ subcase, such a limit exists. It is represented by the vertical red solid line at ${q=0}$ with ${M<0}$, and denoted as RN--BR$_0$. Surprisingly, this limit \emph{is not the same} as the Schwarzschild--BR solution (which has ${r_0=0}$) discussed in Sec.~\ref{sec_r_0_metr}. The Schwarzschild--BR spacetime is represented in our scheme by the vertical green line at ${q=0}$. Notice that in this case, one is able to perform a reflection ${M \leftrightarrow -M}$ which we indicate by the green double-arrow near the origin of the scheme. The fact that in the ${q\to 0}$ limit one does \emph{not} obtain the Schwarzschild--BR solution from the RN--BR$_-$ solution  with a negative mass (as explained also at the end of Sec.~\ref{section-VdBC}) is denoted by two \emph{distinct} parallel lines (green and red) at ${q=0}$, ${M<0}$.

Interestingly, even though the uncharged limit $q\to 0$ of the RN--BR subclass does \emph{not} give the Schwarzschild--BR solution, its limit when \emph{both} ${q\to 0}$ and ${M\to 0}$ exists, leading to the Bertotti--Robinson spacetime (depicted as a purple dot in the center of the scheme). Such a limit was described in Sec.~\ref{sec:m=0,q=0}. The same is true also for the ${m \to 0}$ limit of the uncharged black holes with ${r_0=0}$, see Sec.~\ref{sec_r_0_metr}.

\vspace{4mm}
Let us now discuss how this picture changes when the external electromagnetic field is switched-off, see the upper part of Fig.~\ref{FIG_scheme}. Such a case ${B=0}$ was discussed in Sec.~\ref{sec:B=0}. We showed that the accelerating RN--BR$_\pm$ spacetimes become the Reissner--Nordstr\"{o}m black holes without acceleration. Moreover, in Sec.~\ref{sec:B=0} we argued that for ${B=0}$ the mass parameters $M_{\pm}$ can be redefined so that the \emph{naked singularities} can also be described in this case (this is indicated in our scheme by the extensions of blue and red hatching beyond the line representing the external case into the parameter domains with ${|M|<|q|}$). Notice that for the generic case with $B\neq 0$, described in Sec. \ref{sec:gen}, this extension of domains of the definition of $M_{\pm}$ is not possible, which is why the corresponding domains ${|M|<|q|}$ in the lower part of the scheme are not hatched.

For the Reissner--Nordstr\"{o}m spacetime it is well-known that one can change the signs of \emph{both} the mass $M$ and the charge $q$ parameters. This is indicated by the arrows around the ${q=0}$ and ${M=0}$ axis, swapping the regions to those with \emph{dashed} red and blue hatching. As was already mentioned, this cannot be done in the generic ${B\ne0}$ case, see Sec.~\ref{sec:gen}.

Finally, it can be seen from the upper part of the scheme on Fig.~\ref{FIG_scheme} that in the case ${B=0}$ the limit ${q\to 0}$ exists for \emph{any} sign of the charge $q$, and \emph{any} mass~$M$. It gives the classic Schwarzschild solution, depicted here by the green vertical line at ${q=0}$. This happens because in the  ${B \to 0}$ limit the parameter $r_0$ becomes \emph{redundant}. It can be eliminated from all the metric functions, allowing for the ${q\to 0}$ limit without any restrictions.

The last subcase we wish to mention is the existence of the ${M\to 0}$,~${q\to 0}$ limit that gives the usual Minkowski spacetime, as expected (a purple dot in the center of the upper scheme).

\vspace{4mm}

This summarizes the physical interpretation of all possible subcases of the \emph{non-twisting} spacetimes in the class of \emph{type~D} solutions with a \emph{non-aligned electromagnetic field} \cite{Ovcharenko2025} (equivalent to solutions of the Einstein--Maxwell equations found in \cite{VandenBergh2020}). We presented them in very convenient coordinates and parameterizations. These are interesting \emph{charged black holes of the Reissner--Nordstr\"{o}m type immersed in the Bertotti--Robinson electromagnetic field} (and thus can be abbreviated as RN--BR). In general, they \emph{accelerate} due to the interaction between the charge of the black hole and the external field, and also due to the presence of cosmic strings along the axes of symmetry. All these spacetimes are of algebraic type~D (unlike the Ernst solution \cite{Ernst1976} for charged black holes accelerating in external Melvin electromagnetic field, which is of a general type~I, see Sec.~14.2.1 in \cite{GriffithsPodolsky:2009}). Moreover, it contains several famous black holes as special subcases, namely the charged C-metric and Reissner--Nordstr\"{o}m black holes, as well as the new interesting Schwarzschild--BR solution \cite{Podolsky2025, Ovcharenko2025}. Our next task is the investigation of their rotating generalizations.

\subsection*{Supplementary material}
Main expressions and derivations related to this paper are contained in the supplementary
Wolfram Mathematica file \cite{supp_mat}.

\subsection*{Acknowledgements}

This work was supported by the Czech Science Foundation Grant No.~GA\v{C}R 26-22381S, and by the Charles University Grant No.~GAUK 260325. We thank Roberto Emparan for valuable discussions concerning the extremal subcase.


\begin{thebibliography}{10}

\bibitem{GriffithsPodolsky:2009} J.~B.~Griffiths and J.~Podolsk\'{y}, {\em Exact Space-Times in Einstein's General Relativity}
(Cambridge University Press, Cambridge, 2009).

\bibitem{Stephani2003} H.~Stephani, D.~Kramer, M.~MacCallum, C.~Hoenselaers and E.~Herlt, {\em Exact Solutions of Einstein's Field Equations}
(Cambridge University Press, Cambridge, 2003).

\bibitem{Frolov2017} V.~P.~Frolov, P.~Krtou\v{s} and D.~Kubiz\v{n}\'{a}k, Black holes, hidden symmetries, and complete integrability, Living. Rev. Relativ. {\bf 20} (2017) 6.

\bibitem{Teukolsky1973} S.~A.~Teukolsky, Perturbations of a rotating black hole. I. Fundamental equations for gravitational, electromagnetic, and neutrino-field perturbations, Astrophys. J. {\bf 185} (1973) 635.

\bibitem{Carter1968} B.~Carter, Hamilton--Jacobi and Schrodinger separable solutions of Einstein's equations, Commun. Math. Phys. {\bf 10} (1968) 280--310.

\bibitem{Plebanski1975} J. F. Pleba\'{n}ski, A class of solutions of Einstein--Maxwell equations, Ann. Phys. (N. Y.) {\bf 90} (1975) 196--255.

\bibitem{Plebanski1976} J. F. Pleba\'{n}ski and M. Demia\'{n}ski, Rotating, charged, and uniformly accelerating mass in
general relativity, Ann. Phys. (N. Y.) {\bf 98} (1976) 98--127.

\bibitem{GriffithsPodolsky:2005} J.~B.~Griffiths and J.~Podolsk\'y, Accelerating and rotating black holes, Class.~Quantum Grav. {\bf 22} (2005) 3467--3479.

\bibitem{GriffithsPodolsky:2006} J.~B.~Griffiths and J.~Podolsk\'y, A new look at the Pleba\'nski--Demia\'nski family of solutions, Int.~J. Mod.~Phys.~D {\bf 15} (2006) 335--369.

\bibitem{PodolskyGriffiths:2006} J. Podolsk\'y and J. B. Griffiths, Accelerating Kerr--Newman black holes in (anti-)de Sitter space-time, Phys. Rev.~D {\bf 73} (2006) 044018.

\bibitem{Vratny2021} J.~Podolsk\'{y} and A. Vr\'{a}tn\'{y}, New improved form of black
holes of type D, Phys. Rev. D {\bf 104} (2021) 084078.

\bibitem{Vratny2023} J.~Podolsk\'{y} and A.~Vr\'{a}tn\'{y}, New form of all black holes of
type D with a cosmological constant, Phys. Rev. D {\bf 107} (2023) 084034.

\bibitem{Ovcharenko2025a} H. Ovcharenko, J. Podolsk\'{y} and M. Astorino, Black holes of type D revisited: Relating their various metric forms, Phys. Rev. D {\bf 111} (2025) 024038.

\bibitem{Ovcharenko2025b} H. Ovcharenko, J. Podolsk\'{y} and M. Astorino, Revisiting black holes of algebraic type D with a cosmological constant, Phys. Rev. D {\bf 111} (2025) 084016.

\bibitem{Griffiths2006_2} J.~B.~Griffiths, P.~Krtou\v{s} and J.~Podolsk\'{y}, Interpreting the C-metric, Class. Quantum Gravity {\bf 23} (2006) 6745--6766.

\bibitem{Anabalon2018} A.~Anabal\'{o}n, M.~Appels, R.~Gregory, D.~Kubiz\v{n}\'{a}k, R.~B.~Mann and A.~\"{O}vg\"{u}n, Holographic thermodynamics of accelerating black holes, Phys. Rev. D {\bf 98} (2018) 104038.

\bibitem{Anabalon2019} A.~Anabal\'{o}n, F.~Gray, R.~Gregory, D.~Kubiz\v{n}\'{a}k and R.~B.~Mann, Thermodynamics of charged, rotating, and accelerating black holes, J. High Energ. Phys. {\bf 2019} (2019) 96.

\bibitem{Liu2022} H.-S.~Liu, H.~L\"{u} and L.~Ma, Thermodynamics of Taub-NUT and Plebanski solutions, J. High Energ. Phys. {\bf 2022} (2022) 174.

\bibitem{KolarKrtousOssowski2025} I.~Kol\'{a}\v{r}, P.~Krtou\v{s} and M.~Ossowski, Conical singularity in spacetimes with NUT is observer dependent, Phys. Rev. D {\bf 112} (2025) 104021.

\bibitem{Bonnor1954} W.~B.~Bonnor, Static magnetic fields in general relativity, Proc. Phys. Soc. A {\bf 67} (1954) 225--232.

\bibitem{Melvin1964} M.~A.~Melvin, Pure magnetic and electric geons, Phys. Lett. {\bf 8} (1964) 65.

\bibitem{Ernst1976_1} F.~J.~Ernst, Black holes in a magnetic universe, J. Math. Phys. {\bf 17} (1976) 54.

\bibitem{Ernst1976_2} F.~J.~Ernst and W.~J.~Wild, Kerr black holes in a magnetic universe, J. Math. Phys. {\bf 17} (1976) 182.

\bibitem{Melvin1966} M.~A.~Melvin and J.~S.~Wallingford, Orbits in a magnetic universe, J. Math. Phys. {\bf 7} (1966) 333--340.

\bibitem{Galtsov1976} D.~V.~Galtsov and V.~I.~Petukhov, Black hole in an external magnetic field, Sov. Phys. JETP
{\bf 47} (1978) 419.

\bibitem{Bizyaev2024} I.~Bizyaev, Classification of the trajectories of uncharged particles in the Schwarzschild--Melvin metric, Phys. Rev. D {\bf 110} (2024) 104031.

\bibitem{Dadhich1979} N.~Dadhich, C.~Hoenselaers and C.~V.~Vishveshwara, Trajectories of charged particles in the static Ernst space-time, J. Phys. A: Math. Gen. {\bf 12} (1979) 215--221.

\bibitem{VandenBergh2020} N. Van den Bergh and J. Carminati, Non-aligned Einstein--Maxwell Robinson--Trautman fields of Petrov type D, Class. Quantum Gravity {\bf 37} (2020) 215010.

\bibitem{Ovcharenko2025} H. Ovcharenko and J. Podolsk\'{y}, New class of rotating charged black holes with nonaligned electromagnetic field, Phys. Rev. D {\bf 112} (2025) 064076

\bibitem{Podolsky2025} J. Podolsk\'{y} and H. Ovcharenko, Kerr black hole in a
uniform Bertotti--Robinson magnetic field: An exact solution, Phys. Rev. Lett. {\bf 135} (2025) 181401.

\bibitem{Zeng2025} X.-X.~Zeng and K.~Wang, Energy extraction from the Kerr--Bertotti--Robinson black hole via magnetic reconnection in a circular and a plunging plasma, Phys. Rev. D {\bf 112} (2025) 064032.

\bibitem{Mirkhaydarov2025} M. Mirkhaydarov, T. Xamidov, P. Sheoran, S. Shaymatov and H. Nandan, Non-monotonic enhancement of the magnetic Penrose process in Kerr-Bertotti-Robinson spacetime and its implication for electron acceleration (2025), arXiv:2601.09919 [gr-qc].

\bibitem{Wang2025} X.~Wang, Y.~Hou, X.~Wan, M.~Guo and B.~Chen, Geodesics and shadows in the Kerr--Bertotti--Robinson black hole spacetime (2025), arXiv:2507.22494 [gr-qc].

\bibitem{Zeng2025_2} X.-X. Zeng, C.-Y. Yang and H. Yu, Optical characteristics of the Kerr--Bertotti--Robinson black hole, Eur. Phys. J. C {\bf 85} (2025) 1242.

\bibitem{Ali2025}  H.~Ali and S. G. Ghosh, Parameter estimation of Kerr--Bertotti--Robinson black holes using their shadows, J. Cosmol. Astropart. Phys. {\bf 018} (2026).

\bibitem{Gray2025} F. Gray, D. Kubiz\v{n}\'{a}k, H. Ovcharenko and J. Podolsk\'{y}, Hidden symmetries and separability structures of Ovcharenko--Podolsk\'y and conformal-to-Carter spacetimes, Phys. Rev. D {\bf 113} (2026) in press,  	 arXiv:2511.21538 [gr-qc].

\bibitem{Zhang2025} Y.-K. Zhang and S.-W. Wei, Effects of magnetic fields on spinning test particles orbiting Kerr--Bertotti--Robinson black holes (2025), arXiv:2510.07914 [gr-qc].

\bibitem{Astorino2025} M.~Astorino, Black holes in the external Bertotti--Robinson--Bonnor--Melvin electromagnetic field, Phys. Rev. D {\bf 112} (2025) 104077.

\bibitem{Alekseev1996} G.~A.~Alekseev and A.~A.~Garcia, Schwarzschild black hole immersed in a homogeneous electromagnetic field, Phys. Rev. D {\bf 53} (1996) 1853.

\bibitem{Alekseev2025} G.~A.~Alekseev, Charged black hole accelerated by spatially homogeneous electric field of Bertotti--Robinson ($AdS^2\times S^2$) space-time, (2025) [arXiv: 2511.06082].

\bibitem{supp_mat} Hryhorii Ovcharenko, https://doi.org/10.5281/zenodo.18619443 (2026).

\bibitem{Ortaggio2018} M.~Ortaggio and M.~Astorino, Ultrarelativistic boost of a black hole in the magnetic universe of Levi-Civita--Bertotti--Robinson, Phys. Rev. D {\bf 97} (2018) 104052.

\bibitem{Ortaggio2025} M.~Ortaggio, Einstein--Maxwell fields as solutions of Einstein gravity coupled to conformally invariant non-linear electrodynamics, (2025) [arXiv:2511.13665].

\bibitem{Alekseev1985} G.~A.~Alekseev, The method of the inverse problem of scattering and the singular integral equations for interacting massless fields, Soviet Phys. Dokl. {\bf 30} (1985) 565--568.

\bibitem{Alekseev1988} G.~A.~Alekseev, Exact solutions in the general theory of relativity, Proc. Steklov Inst. Math. {\bf 176} (1988) 215--262.

\bibitem{Alekseev2000} G.~A.~Alekseev and J.~B.~Griffiths, Infinite hierarchies of exact solutions of the Einstein and Einstein--Maxwell equations for interacting waves and inhomogeneous cosmologies, Phys. Rev. Lett. {\bf 84} (2000), 5247.

\bibitem{Alekseev2001} G.~A.~Alekseev and J.~B.~Griffiths, Solving the characteristic initial-value problem for colliding plane gravitational and electromagnetic waves, Phys. Rev. Lett. {\bf 87} (2001) 221101.

\bibitem{Alekseev2004} G.~A.~Alekseev and J.~B.~Griffiths, Collision of plane gravitational and electromagnetic waves in a Minkowski background: solution of the characteristic initial value problem, Class. Quantum Grav. {\bf 21} (2004) 5623-5654.

\bibitem{Halilsoy1993} M.~Halilsoy, Interpolation of the Schwarzschild and Bertotti--Robinson solutions,  Gen. Relat. Gravit. {\bf 25} (1993) 275--280.

\bibitem{Ernst1976} F.~J.~Ernst, Removal of the nodal singularity of the C-metric, J. Math. Phys. {\bf 17} (1976) 515--516.


\end{thebibliography}
\end{document}